\documentclass[sigconf]{acmart}
\AtBeginDocument{%
  }

\copyrightyear{2026}
\acmYear{2026}
\setcopyright{cc}
\setcctype{by}
\acmConference[CHI '26]{Proceedings of the 2026 CHI Conference on Human Factors in Computing Systems}{April 13--17, 2026}{Barcelona, Spain}
\acmBooktitle{Proceedings of the 2026 CHI Conference on Human Factors in Computing Systems (CHI '26), April 13--17, 2026, Barcelona, Spain}
\acmPrice{}
\acmDOI{10.1145/3772318.3790287}
\acmISBN{979-8-4007-2278-3/2026/04}

\newcommand{\inlinefig}[1]{%
  \raisebox{-0.2\height}{\includegraphics[height=1em]{#1}}%
}

\usepackage{soul}
\definecolor{hlc}{HTML}{FFFFB3}  
\sethlcolor{hlc}

\begin{document}

\title{Studying the Separability of Visual Channel Pairs in Symbol Maps}

\author{Poorna Talkad Sukumar}
\affiliation{%
  \institution{Tandon School of Engineering\\ New York University}
  \city{Brooklyn}
  \state{New York}
  \country{USA}
}
\email{poorna.t.s@gmail.com}

\author{Maurizio Porfiri}
\affiliation{%
  \institution{Tandon School of Engineering\\ New York University}
  \city{Brooklyn}
  \state{New York}
  \country{USA}
}
\email{mporfiri@nyu.edu}

\author{Oded Nov}
\affiliation{%
  \institution{Tandon School of Engineering\\ New York University}
  \city{Brooklyn}
  \state{New York}
  \country{USA}
}
\email{on272@nyu.edu}

\renewcommand{\shortauthors}{Talkad Sukumar et al.}

\begin{abstract}
  Visualizations often encode multivariate data by mapping attributes to distinct visual channels such as color, size, or shape. The effectiveness of these encodings depends on \emph{separability}—the extent to which channels can be perceived independently. Yet systematic evidence for separability, especially in map-based contexts, is lacking. We present a crowdsourced experiment that evaluates the separability of four channel pairs—color (ordered) × shape, color (ordered) × size, size × shape, and size × orientation—in the context of bivariate symbol maps. Both accuracy and speed analyses show that color × shape is the most separable and size × orientation the least separable, while size × color and size × shape do not differ. Separability also proved asymmetric—performance depended on which channel encoded the task-relevant variable, with color and shape outperforming size, and square shape especially difficult to discriminate.  Our findings advance the empirical understanding of visual separability, with implications for multivariate map design.

\end{abstract}


\begin{teaserfigure}
  \includegraphics[width=\textwidth]{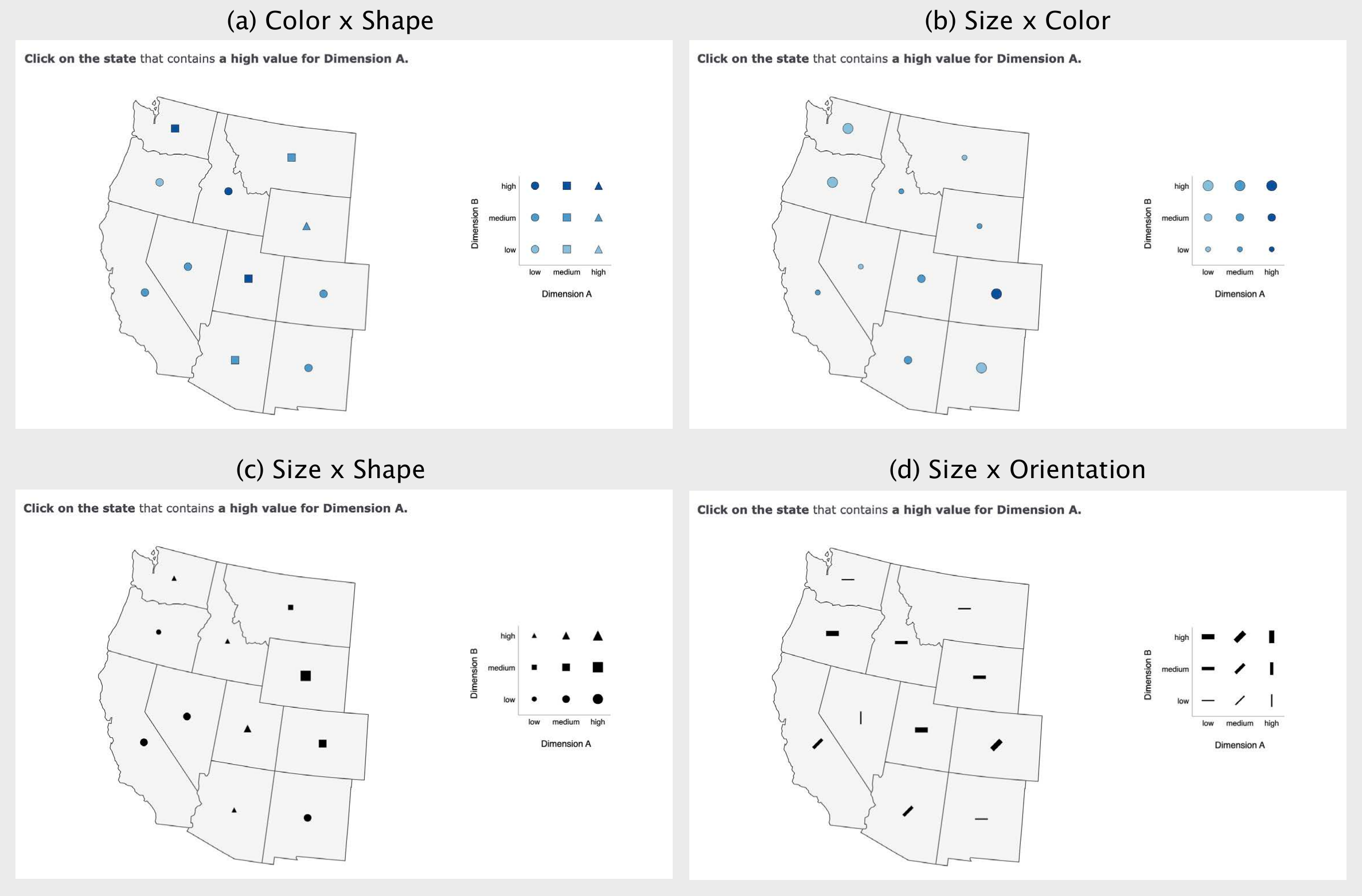}
  \caption{Example stimuli for the four visual channel pairs tested: (a) Color × Shape, (b) Size × Color, (c) Size × Shape, and (d) Size × Orientation. Each consists of a symbol map of the western U.S. where each state is randomly assigned a bivariate symbol, with one visual channel encoding Dimension A (task-relevant) and the other encoding Dimension B. Each channel had three levels (low, medium, high), producing a 3 × 3 symbol matrix shown in the legend. On each trial, participants were asked to click on the state containing a high value of Dimension A. Each trial contained exactly one high-value instance of Dimension A. Across trials, the specific high-value level of the Dimension A channel was varied, and participants also completed flipped versions where Dimensions A and B were swapped. 
}
  \Description{Example Stimuli for the Four Visual Channel Pairs. This figure shows four example bivariate symbol maps of the western United States. Each map contains 11 states, each with a symbol at its centroid. The four panels illustrate the visual channel pairs: Color × Shape, Size × Color, Size × Shape, and Size × Orientation. Each map includes a legend showing a 3×3 grid of symbol combinations, with Dimension A levels on one axis and Dimension B levels on the other. Participants clicked the state containing the high-value symbol for Dimension A.}
  \label{fig:teaser}
\end{teaserfigure}


\maketitle

\section{Introduction}
Visualizations often rely on the use of multiple visual channels, such as color, size, or shape, to encode multivariate data. Separability refers to the extent to which these different encoding channels can be perceived independently of one another \cite [Ch. 5]{munzner2025visualization}. Visual channels lie on a continuum from separable to integral: at one end, separable pairs (e.g., spatial position and color) allow users to more easily attend to each encoded attribute independently without interference; at the other end, integral pairs (e.g., horizontal and vertical size) blend perceptually into a single, inseparable composite such as area. For example, when size is combined with color, the larger symbols often dominate attention, making it harder to distinguish colors in smaller symbols, illustrating that size is not fully separable from color \cite{szafir2017modeling, stone2014engineering}. Such interference effects can compromise a viewer’s ability to accurately interpret multivariate visualizations, particularly when tasks require isolating one dimension of information from another. 

The integral–separable continuum is foundational in perception theory and serves as a simple design guideline for visualization: separable pairs are typically preferred for encoding multiple independent data attributes, while integral pairs are used to redundantly encode the same attribute \cite[Ch. 5]{ware2019information}. Although this continuum has been discussed for decades \cite{garner1974processing}, the placement of visual channel pairs along it has been shaped largely by theoretical assumptions and expert heuristics rather than systematic empirical testing. As Ware \cite[Ch. 5]{ware2019information} notes, the integral–separable distinction itself is theoretically simplistic—lacking a mechanistic basis and overlooking many experimentally observed exceptions and asymmetries.

In addition, the most frequently cited continuum excludes several important dimensions, such as ordered color and shape × size, leaving the framework incomplete. Some studies have begun to address these limitations by empirically examining separability, often focusing on scatterplots \cite{smart2019measuring}. However, a critical gap remains in understanding how separability operates in spatially embedded visualizations, such as maps, where spatial context and symbol interactions may introduce different forms of interference.

To advance empirical understanding in this space, we conducted a preregistered\footnote{\url{https://osf.io/sf5wm/?view_only=5381119f126645eb99e0deaa53587392}} within-subjects experiment that systematically tested separability across four visual channel pairs—\inlinefig{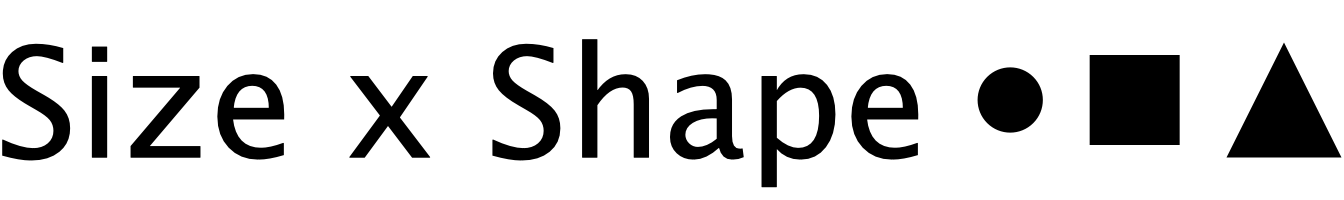}, \inlinefig{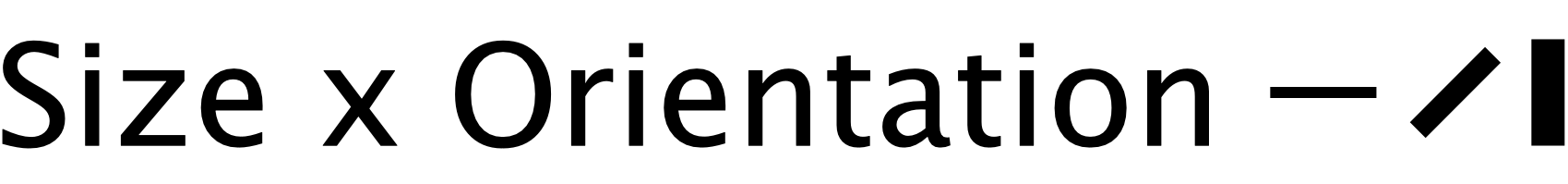}, \inlinefig{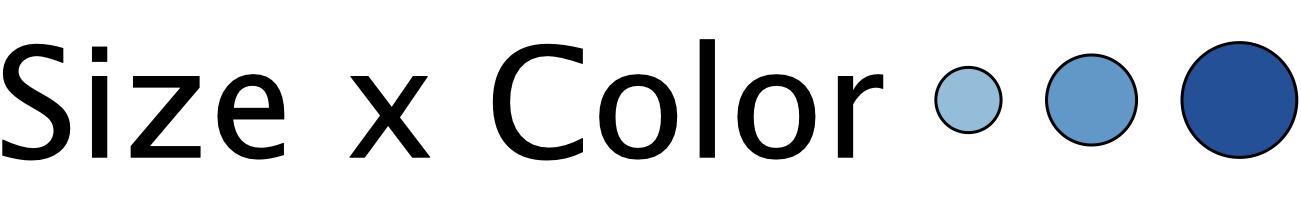}, and \inlinefig{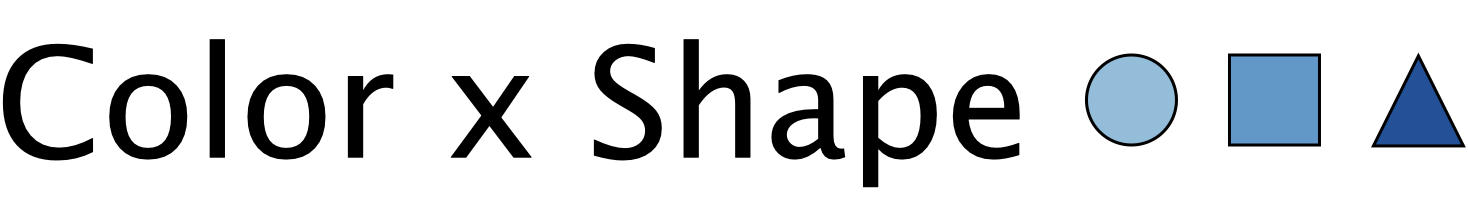}— 

\noindent within symbol map visualizations. A \textit{symbol map} is a map-based visualization in which symbols (e.g., circles, squares, or icons) are placed at geographic locations to represent data values, with visual attributes such as size, color, or shape used to encode quantitative or categorical information. Drawing inspiration from Dobson’s early map perception studies \cite{dobson1983visual} and MacEachren’s illustrative examples of visual variables in map design \cite[Ch. 3]{maceachren2004maps}, we used bivariate symbol maps of 11 western U.S. states. Each state was represented by a symbol encoding two dimensions of information, each varying across three levels (low, medium, and high). In each trial, participants identified the state with the highest value of a task-relevant dimension, using a reference legend. We also flipped which visual channel encoded the task-relevant variable to examine how channel assignment affected performance, as differences across flips can indicate asymmetries in perceptual interference.

Our findings provide an empirically grounded evaluation of visual channel separability in symbol map visualizations. Both accuracy and response time analyses converged in identifying \inlinefig{color_shape.png} as the strongest pairing—yielding high accuracy and fast responses—and \inlinefig{size_orientation.png} as the weakest, with consistently low accuracy and slow responses. However, performance was not strictly linear: the two mid-ranking pairs show a trade-off—\inlinefig{size_shape.png} is preferable if accuracy matters more than speed, and vice versa for \inlinefig{size_color.png}. We also found that separability was asymmetric, with outcomes depending on which channel encoded the task-relevant variable, and that square shape was especially challenging to discriminate. Together, these results highlight that separability depends not only on channel pairing but also on channel assignment and value distribution of the interfering channel, underscoring the need to move beyond pairwise orderings toward a more nuanced understanding of separability in visualization design.

\textbf{Contributions:} This study advances empirical understanding of separability in map-based visualizations. We present a systematic evaluation of four visual channel pairs in the context of symbol maps. Our findings reveal differences in task performance and perceptual asymmetries across channel pairings and assignments, underscoring the importance of considering both the combination of channels and the assignment of variables to channels in multivariate designs. 
Together, these results inform design choices for bivariate symbol maps and contribute to a broader theoretical understanding of the integral–separable continuum.

\section{Background}

\subsection{The Integral–Separable Continuum}

The concept of integral and separable dimensions originates in Garner’s seminal work on multidimensional perception \cite{garner1974processing}. Two attributes are considered integral when they are perceived holistically and cannot be easily disentangled. For example, the width and height of a rectangle fuse into an overall perception of shape or area. By contrast, separable dimensions can be judged independently, such as position and color of a symbol, which can be attended to in isolation without substantial interference. Rather than a strict dichotomy, later accounts emphasize a continuum \cite{shortridge1982stimulus}: even highly separable pairs show some interaction, and integral pairs can sometimes be parsed analytically with effort. 

Integral–separable distinctions have been studied through a set of well-established tasks, usually limited to pairs of visual channels. In \textit{restricted classification tasks}, participants are asked to group stimuli; integral dimensions lead to holistic groupings, while separable dimensions promote dimension-specific matches \cite[Ch.~5]{ware2019information}. In \textit{speeded classification tasks}, observers classify items along one attribute while another varies; integral dimensions produce strong interference, whereas separable ones show little interference. Finally, \textit{redundant coding tasks} examine performance when the same variable is encoded with two dimensions, often yielding benefits for integral pairs but little gain for separable ones. Hence, the integral-separable continuum has become a key design heuristic in information visualization: separable pairs are recommended for encoding distinct variables, while integral pairs are better suited for redundant encodings that reinforce the same variable \cite[Ch.~5]{ware2019information}.

Current expert-based rankings of this continuum place certain channel pairs toward the integral end, such as opponent color channels (red–green and yellow–blue) or the x– and y–size of objects, which tend to be perceived holistically. Pairs such as size and orientation fall closer to the middle, showing partial independence but also interference. Toward the separable end, combinations such as color with shape, size, or orientation, or group location with color, are judged to be more easily distinguished \cite[Ch.5]{ware2019information}. These rankings, form the basis for our study hypothesis about which pairs will show greater or lesser separability in symbol maps.

Most research examines separability in two-channel combinations, yet real-world graphics often employ three or more encodings. Research cautions that interference can compound with additional channels: dimensions that appear separable in pairs may produce unexpected dependencies in higher-dimensional settings \cite[Ch.~3]{maceachren2004maps}.

Separability is closely related to, but distinct from, other low-level perceptual phenomena such as preattentive processing and visual pop-out. Preattentive processing refers to the rapid, parallel detection of basic visual features such as hue, orientation, size, or motion, typically within 200–250 ms, before focused attention is engaged \cite{healey2011attention}. Pop-out is the search effect that results when a target differs strongly on such a feature, allowing it to be spotted quickly regardless of the number of distractors — for example, a red circle among blue ones. Separability, by contrast, concerns not the speed of detection but the perceptual independence of multiple channels used in combination. Two channels may each support popout when varied alone, but interfere when combined.

The integral–separable continuum provides a valuable but incomplete framework  \cite[Ch.~5]{ware2019information}. Existing rankings along this continuum are largely based on expert heuristics and design assumptions rather than systematic empirical testing. As a result, current guidelines are often simplistic: they offer general predictions about pairwise combinations without accounting for task-dependent performance, or asymmetries in how one channel may interfere with another. In particular, there is a lack of fine-grained evidence for specific channel interactions, especially in multivariate and spatially embedded contexts such as maps. To advance beyond these limitations, this study undertakes a systematic empirical investigation of separability across visual channel pairs in symbol maps.

\subsection{Graphical Perception}

Graphical perception research investigates how effectively visual channels support data interpretation tasks. Foundational work by Cleveland and McGill \cite{cleveland1984graphical} established a ranking of channels for quantitative judgments, placing position at the top, followed by length, angle, and area. These findings have been replicated and extended using large-scale online experiments \cite{heer2010crowdsourcing} and across diverse visualization types such as bar charts, scatterplots, and pie charts. More recent studies emphasize that perceptual effectiveness is not fixed: factors such as task type, data distribution, and visualization context strongly mediate performance \cite{kim2018assessing}.

Synthesizing this literature, Zeng and Battle \cite{zeng2023review} reviewed 59 papers across ten analysis tasks, showing that while position consistently supports accurate judgments, other encodings such as color, size, and shape exhibit task- and context-dependent performance. For example, color saturation can effectively encode quantitative data, while color hue and shape are better suited for categorical distinctions. However, their review also highlights a gap: although many studies establish rankings of individual channels, far fewer examine interference effects—how one channel influences judgments made with another.

Perception-based visualization studies also differ in the tasks they employ, which in turn shape the perceptual mechanisms being engaged and the kinds of interference revealed. Identification tasks emphasize local discriminability, while comparison, ranking, or trend-detection tasks probe higher-level integration across stimuli. A recent survey of perception-based visualization studies highlights this diversity, showing that different tasks can lead to distinct findings and methodological emphases \cite{quadri2021survey}.

The channels most relevant to our study—shape, color, and size—have been investigated individually. Shape has primarily been studied for categorical discrimination, where filled versus unfilled or open versus closed marks produce different discriminability and interference patterns \cite{burlinson2017open, smart2019measuring}. Color encodings, long known to be sensitive to display conditions and background \cite{brainard1992asymmetric, oicherman2008effect, stone2014engineering, szafir2017modeling}, can represent both qualitative and quantitative data when perceptually uniform scales such as ColorBrewer or CIELAB interpolation are used. Size encodings, typically represented by area, are less precise than position or length but remain common in practice; moreover, size perception itself can depend on shape, with squares judged most accurately, circles intermediate, and triangles least accurately \cite{Qi2022}. 

Scatterplots provide another context where perception has been extensively studied. Tseng et al. \cite{tseng2023measuring} showed that categorical perception in color-coded scatterplots depends on both the number of categories and the chosen color palette, with accuracy degrading as categories increase. Lu et al. \cite{Lu2023} introduced interactive color highlighting to preserve context while maintaining class discriminability, addressing limitations of conventional dimming. Beyond color, Tseng et al. \cite{tsengshape} demonstrated that shape palettes vary widely in effectiveness, with empirically derived models outperforming heuristic design choices. Liu et al. \cite{Liu_orientation} further extended scatterplot perception research by showing that mark orientation can systematically bias or guide regression-by-eye judgments, particularly when trends are ambiguous due to outliers or dispersion.

Despite this substantial body of work, Zeng and Battle note that channels such as shape and orientation remain comparatively underexplored, and that multi-encoding interactions are rarely examined \cite{zeng2023review}. Our study addresses this gap by empirically testing the separability of under-studied encodings (shape and orientation) and their interactions with other visual channels in symbol maps.

\subsection{Empirical Studies on Separability in Multivariate Visualizations}

Prior work by Smart and Szafir \cite{smart2019measuring} systematically measured how shape, size, and color interact in scatterplots. They found that separability among these channels is asymmetric: shape more strongly influences size and color perception than the reverse, with filled or denser shapes perceived as larger and making colors easier to discriminate, whereas size and color had only limited effects on shape discrimination. Although their experiments focused on scatterplots, symbol maps similarly rely on small, discrete marks where interference between visual channels can affect task performance. We therefore treat their findings as evidence of a broader perceptual principle—that shape provides a more robust basis for attending to a target dimension than size or color—which motivates our directional hypotheses regarding asymmetries in \text{Size} $\times$ \text{Shape} and \text{Color} $\times$ \text{Shape} pairings.

In cartography, Dobson’s classic experiments on multivariate symbol maps \cite{dobson1983visual}, provide a key methodological precedent. Dobson presented participants with maps of the western United States containing graduated circles that encoded economic data. Subjects performed location, categorization, and comparative judgment tasks to evaluate how effectively map readers could extract information under different symbolization strategies. Dobson found that pairing size with color value improved performance over size alone, suggesting that redundant encodings based on partially integral channels could facilitate processing.

Our experiment builds on this paradigm by also using maps of western U.S. states with bivariate symbols, but it differs in two key respects. First, we employ a single identification task, requiring participants to click on the state corresponding to a target value, rather than multiple task types. Second, instead of testing whether redundancy improves performance, we focus on how different pairs of visual channels vary in their degree of separability or interference.

\section{Experiment}

\subsection{Study Overview and Design}

We conducted a preregistered\footnote{\url{https://osf.io/sf5wm/?view_only=5381119f126645eb99e0deaa53587392}} within-subjects repeated-measures experiment to evaluate the perceptual separability of visual channel pairs in the context of bivariate symbol maps. The study examined how different combinations of visual channels affect participants’ ability to read and interpret such maps. Participants viewed maps of the western United States overlaid with symbols that simultaneously encoded two data dimensions: Dimension A (task-relevant) and Dimension B (task-irrelevant). The key objective was to assess how effectively participants could isolate and interpret Dimension A while ignoring interference from Dimension B. To this end, they performed a forced-choice identification task, selecting the state containing a symbol that represented a high value on Dimension A.

Our task design follows the established goals of separability research, which focus on measuring “the ability to attend to one dimension and ignore another” \cite [Ch. 3]{maceachren2004maps} . This selective-attention paradigm originates in Garner’s and Pomerantz’s \cite{pomerantz1973stimules} framework and is typically evaluated through simplified perceptual tasks that isolate a single visual dimension. Based on the categorization task (identifying the category for a specific state) used in Dobson’s separability experiments \cite{dobson1983visual}, we use a reverse-categorization task (clicking on a state that corresponds to a given category). This task does not require prior geographic knowledge and ensures that performance depends only on perceptual discrimination of the symbol encodings. 

As with most controlled graphical-perception experiments, our approach also involves an inherent trade-off between task simplicity and ecological realism: simpler tasks sacrifice real-world complexity but allow far greater experimental control, enabling clean interpretation of cross-channel interference and selective-attention effects. This approach aligns with contemporary work: Smart and Szafir \cite{smart2019measuring}, for example, evaluated the separability of color, shape, and size using a deliberately simple binary-comparison task—judging which of two marks, with fixed distance between them in all stimuli, differed along a given dimension. Their work, like ours, employs a low-level perceptual task rather than complex analytical activities such as correlation judgment. These simplified designs reflect a methodological consensus: isolating selective attention under controlled conditions provides the clearest diagnostic evidence of separability.

The perceptual mechanisms we study—selective attention and interference—are directly applicable to real-world cartographic reasoning. For example, analysts interpreting bivariate maps of rainfall and elevation, or journalists visualizing unemployment rate and education level by state, often must attend selectively to one variable while ignoring another \cite[Ch.~3]{maceachren2004maps}. Such everyday analytic activities build on the same selective-attention processes assessed in separability experiments.  Simplified tasks therefore serve as diagnostic foundations for understanding how map readers integrate multiple variables in real-world analyses \cite[Ch.~3]{maceachren2004maps}. While our study focuses on perceptual separability, we agree that examining higher-level analytical tasks—such as comparison or correlation assessment—represents an important future direction.

We tested four visual channel pairs commonly used in multivariate map visualizations: color × shape, size × color, size × shape, and size × orientation. Each visual channel pair is tested under two flipped conditions to assess potential asymmetries in perception depending on which channel encodes the task-relevant variable (e.g., color encoding Dimension A vs. size encoding Dimension A). 
Example stimuli for each channel pair are shown in Figure \ref{fig:teaser}.

Our choice of visual channel pairs was guided by methodological precedents in Dobson’s work \cite{dobson1983visual} and MacEachren’s book \cite [Ch. 3]{maceachren2004maps}, which commonly illustrate size × orientation, size × shape, and size × color pairings in bivariate symbol maps. We adopted these established combinations—with a minor adaptation in size × shape where we used a square rather than a line to maintain geometric comparability—and added one extension (color × shape) to include a widely used, highly separable pairing. We acknowledge that many other combinations, such as color × orientation, remain to be explored. 
We did not include color × orientation because orientation in our stimuli was encoded using narrow line marks with limited filled surface area relative to the 2D shapes used elsewhere. At the small symbol sizes required for symbol maps, this would have introduced an additional confound—varying effective color area across channel pairs—making it difficult to attribute performance differences specifically to channel interaction rather than mark geometry.

The full design comprised 4 channel pairs × 2 flips × 3 target variations = 24 experimental trials per participant. Trial order and symbol placements were randomized to minimize order and location biases. This repeated-measures design ensured that each participant contributed data across all conditions, increasing sensitivity to subtle differences while controlling for individual variation in baseline performance. We also included two practice trials at the start of the study to familiarize participants with the task, and two attention checks were inserted mid-block to ensure data quality.

\subsection{Hypotheses}

We preregistered the following hypotheses, operationalizing ``better performance'' as \emph{higher accuracy} and \emph{faster response times (RTs) on correct trials}.

\begin{itemize}
    \item \textbf{H1 (Directional; separability across pairs).} There will be a main effect of visual channel pair on performance. Expected order (most to least separable): 
\[
\text{Color} \times \text{Shape} \;>\; \text{Size} \times \text{Color} \;>\; \text{Size} \times \text{Shape} \;>\; \text{Size} \times \text{Orientation}.
\]
\emph{Rationale.} This predicted ordering is based on existing integrality–separability continua \cite [Ch. 5]{ware2019information}. \\

\item \textbf{H2 (Directional; asymmetry in Shape $\times$ Size).} 
Participants will perform better when \textit{Shape} encodes the task-relevant dimension (A) than when \textit{Size} encodes A.  \hfill \break
\emph{Rationale.} As noted by Smart and Szafir \cite{smart2019measuring}, shape systematically biases size perception, whereas size has only weak effects on shape discrimination, suggesting that shape should be more robust when used as the task-relevant channel. \\

\item \textbf{H3 (Directional; asymmetry in Color $\times$ Shape).} 
Participants will perform better when \textit{Shape} encodes A than when \textit{Color} encodes A.  \hfill \break
\emph{Rationale.} Smart and Szafir \cite{smart2019measuring}  also showed that shape strongly influences color discriminability while color has only a limited effect on shape perception. This supports the prediction that shape is the more robust channel in color–shape pairings.  \\

\item \textbf{H4 (Exploratory; asymmetry in remaining pairs).} For \textit{Color} $\times$ \textit{Size} and \textit{Size} $\times$ \textit{Orientation}, we explore potential asymmetries without directional predictions. \hfill \break
\emph{Rationale.} Prior evidence on interference patterns for these pairings, especially in spatially embedded symbol maps, is limited; we therefore treat directionality as an open question. \\

\item \textbf{H5 (Exploratory; influence of target value).}
As an additional, non-preregistered exploratory analysis, we examine how performance varies across target value levels (low, medium, high) of the interfering channel within flips, to provide finer-grained insight into when interference effects are most pronounced. \\
\end{itemize}

Each hypothesis was evaluated using both accuracy and response times (restricted to correct trials), with repeated-measures models applied since all participants completed every condition. Post-hoc pairwise comparisons were adjusted using standard corrections (e.g., Tukey/Holm) to control for inflated Type I error due to multiple testing.


\subsection{Stimuli}

\subsubsection{Map substrate}
Each trial displayed a symbol map of 11 western U.S. states (Washington, Oregon, Idaho, Montana, Wyoming, California, Nevada, Arizona, Utah, Colorado, and New Mexico). A single bivariate symbol was placed at the centroid of each state. To ensure comparability across conditions, all map elements—including the display size and layout of the map (700 × 600 px) and legend—were held constant across trials, with only the overlaid symbols varying by condition. The use of the western U.S. map with 11 states, inspired by Dobson’s study \cite{dobson1983visual}, balances visual search demands by providing enough items to require focused attention while keeping task completion feasible for an online experiment.

\subsubsection{Bivariate symbols and legend} On every trial, each state’s symbol encoded two dimensions (A, B) using one of the four tested visual channel pairs. For all channel pairs we defined three ordered levels per channel (low, medium, high), yielding a 3×3 = 9-symbol vocabulary. 
Our design closely follows the methodological precedent established by Dobson \cite{dobson1983visual} and the examples presented by MacEachren \cite{maceachren2004maps}, where the same western U.S. layout (11 states) and three discrete symbol levels were used to examine interference between bivariate encodings. As in those classic studies, the simplified geographic substrate and restricted number of levels were intentional: they ensure that differences in accuracy and response time can be attributed to visual–variable interaction, rather than to geographic complexity, data density, or crowding effects. Consistent with other controlled graphical-perception work \cite{smart2019measuring}, our goal was to isolate core channel interactions under tightly controlled conditions that maximize internal validity.

Using three levels offered a practical balance--two levels would under-sample potential graded interference effects, while four or more would substantially increase the symbol vocabulary ($\geq$ 16 bivariate combinations) and inflate search time and error rates in an online setting.
A matrix legend showed all nine combinations; axis labels indicated which channel mapped to A vs. B on that trial. The legend supports direct, exemplar-based matching and ensures that participants can perform the task based on appearance rather than needing to infer a numeric scale. 

In each trial, one state was randomly assigned one of the three “high” values of Dimension A, while the remaining states were filled with randomly sampled combinations of the other six symbols. This guaranteed a single correct answer per trial. Because 
fully balancing the ten non-target states across all possible symbol combinations would 
require an infeasibly large number of unique map configurations and introduce repetition 
effects, we randomized the assignment of non-target symbols. To balance representation across all high-value symbols, we constructed three trials for each of the eight visual encoding conditions (4 channel pairs × 2 flips). In each case, a different high-value symbol was assigned to a randomly chosen state, ensuring that all target symbols were tested equally and without ambiguity.

\subsubsection{Channel implementations and level choices}
We used three perceptually distinguishable levels for each visual channel: \\

\noindent\textbf{Color (ordered):} We constructed a three-step sequential blue scale (\#89bedc, \#4e9acb, \#08519c) inspired by the perceptual spacing of the ColorBrewer 3-class Blues ramp \cite{harrower2003colorbrewer}. The hex values were selected to produce approximately uniform and clearly distinguishable steps in lightness, with CIELAB differences of $\Delta E \approx 17.6$ (light–medium) and $\Delta E \approx 34.9$ (medium–dark), ensuring adequate separation even when rendered as small map symbols. This approach follows prior recommendations that color palettes for visualization should be chosen based on perceptual differences rather than raw RGB values, as perceived discriminability decreases with symbol size \cite{stone2014engineering, szafir2017modeling}. In addition, darker tones were used to maintain contrast against the light gray map background. All colored symbols were rendered with a 0.5px black outline.

Color discriminability varies substantially with context. While classical CIELAB work defines $\Delta E \approx 2$--$3$ as a minimal just noticeable difference (JND) threshold under ideal viewing conditions, visualization research shows that considerably larger separations are required for reliable discrimination of small marks embedded in complex scenes. Prior work suggests that $\Delta E$ values of $\sim 6$ for larger marks and $\Delta E \ge 10$--$12$ for small marks are needed for robust separability \cite{szafir2017modeling, stone2014engineering}. Our selected palette provides $\Delta E$ separations of 17.6 and 34.9, which well exceed these recommendations.

We also chose a blue sequential scale rather than green or categorical hues to maintain perceptual monotonicity and ensure accessibility for participants with color blindness. Blue scales are more robust to red–green color vision deficiencies—the most common form of color blindness—and this consideration is particularly relevant because we rely on participants’ self-reported color-blindness status. \\

\textbf{Size:} We used three size levels, 2, 5, and 8 pixels, chosen to approximate a perceptually graded progression. 
For the size × orientation condition, size was manipulated through line stroke-widths of 2px, 5px, and 8px, with line length fixed at 20px. For the size × shape condition, circles (radii = 2, 5, 8px), squares (half-sides = 2, 5, 8px), and triangles (half-heights = 2, 5, 8px) were scaled so that symbols at each size level had the same overall height. This standardization ensured that perceived differences arose from the intended size manipulation rather than unintended differences in shape geometry. 

Perceived symbol size follows a compressive power law (exponent $\approx 0.7$ \cite{stevens2017psychophysics}), meaning that equal physical increments in area do not yield equal perceptual increments. There is no fixed JND for symbol area, as discriminability depends on factors such as separation distance and background clutter \cite{lu2021modeling}. Prior magnitude-scaling work, however, provides a useful reference point: for one circle to appear twice as large as another, its area must be approximately $2.7\times$ larger and its diameter about $1.6\times$ larger \cite{stevens1963subjective}. Although this is not a JND, it illustrates the substantial physical differences required for reliable area-based comparisons. 

Our chosen half-heights (2 px, 5 px, 8 px) produce area ratios of $1:6.25:16$, yielding large physical separations between size levels. While these ratios are not intended to correspond to a specific perceptual threshold, they ensure that the size levels are clearly distinguishable within the spatial constraints of symbol maps. The conditions included in the supplementary material show the size encodings more clearly.
Our chosen sizes keep symbols small enough to avoid overlap within the map context, while being discriminable. \\ 

\textbf{Shape:} We used three simple filled shapes: circle, square, and triangle. Although shape is typically categorical, we applied it to encode an ordinal dimension (low–medium–high), following cartographic precedents such as MacEachren’s bivariate maps where shape was combined with size to represent ordinal data \cite [Fig. 3.40]{maceachren2004maps}. 
In the size × shape condition, these shapes were rendered in black at three size levels (2, 5, 8px). In the color × shape condition, size was held constant at the medium level (5px), and the shapes were instead filled with the three color levels described above. \\

\textbf{Orientation:} We used line segments oriented at 0°, 45°, and 90°. This choice was also inspired by an example presented by MacEachren \cite [Fig. 3.36]{maceachren2004maps}, where orientation was used in ordered bivariate symbol schemes.

\subsection{Procedure}

The study was delivered online via Qualtrics, where visualization stimuli were generated using D3.js implemented through Qualtrics’ embedded JavaScript. To standardize viewing conditions and avoid touch-interaction ambiguities, only desktop and laptop participation was permitted; mobile devices were not allowed.

After providing informed consent and confirming that they had normal or corrected-to-normal vision, participants were introduced to the practice trials. For both the practice block and all subsequent main trials, the task was the same: participants were instructed to click the state containing a symbol with a high value for Dimension A, as indicated in the legend. In the practice block, they completed two trials using the same types of visual encodings as in the main experiment (color × shape and size × orientation), but with alternate symbol levels not used in the main study (color-filled $\lozenge$ instead of $\square$, and orientations at –45°, 0°, and 45° instead of 0°, 45°, and 90°). Feedback was provided after each practice trial, indicating the correct state. These practice trials served solely for familiarization and were excluded from analysis.

The main block consisted of 24 trials (4 visual channel pairs × 2 flips × 3 symbol variations), presented in a randomized order using Qualtrics’ randomization feature. We did not impose an explicit time limit during the study; however, participants were instructed to respond both quickly and accurately.  Response times (ms) were recorded from stimulus onset to the participant’s click response, and both accuracy and response times were captured via Qualtrics Embedded Data. After a click, the study advanced immediately to the next trial. We also included two attention check questions during the main block. Participants who failed either check were excluded.

After the main block, participants completed a demographic questionnaire and reported their familiarity with data visualizations before finishing the study with a short debrief. The Qualtrics survey and visualization conditions are included in the supplementary material.

\subsection{Participants}
Participants were recruited through the Prolific platform. Eligibility criteria required participants to be fluent in English and to complete the study on a desktop or laptop. All participants were from the United States and had
an approval rating of 99\% or greater. Participants with self-reported vision problems or color blindness  were excluded.
We aimed to detect small-to-moderate within-subject effects (Cohen’s $f = 0.15$) with 80\% power at $\alpha = 0.05$. An a priori power analysis using G*Power (repeated-measures ANOVA, within factors, 8 conditions) indicated a required sample size of approximately 180 participants, assuming conservative sphericity corrections ($\epsilon = 0.75$). To allow for exclusions and ensure adequate power, we recruited 200 participants.  
Participants were compensated at a rate of \$2.00 for approximately 8–10 minutes of participation. Median time was 6.37 minutes.

\subsection{Ethics Statement}

This study involved human participants recruited via Prolific. All procedures were reviewed and approved by the university's Institutional Review Board (IRB). Participants provided informed consent prior to participation and were free to withdraw at any time. No personally identifying information was collected beyond platform IDs required for compensation, and data were analyzed only in aggregate to ensure confidentiality. Compensation rates met or exceeded fair payment guidelines.

\section{Results}

We excluded 3 of 200 participants who answered either attention check question incorrectly, as well as those who reported that they were or could be color blind. We excluded trials with response times (RTs) below 300 ms (possible anticipatory or accidental responses) or above 10,000 ms (possible task disengagement or distraction) to remove implausible values and reduce noise in the data. For RT analyses, we included only correctly answered trials to avoid confounding speed with accuracy, and applied a log transformation to correct for the positive skew typical of RT data and to meet the normality assumptions of parametric tests.

\begin{figure*}[h]
  \centering
  \includegraphics[width=1\linewidth]{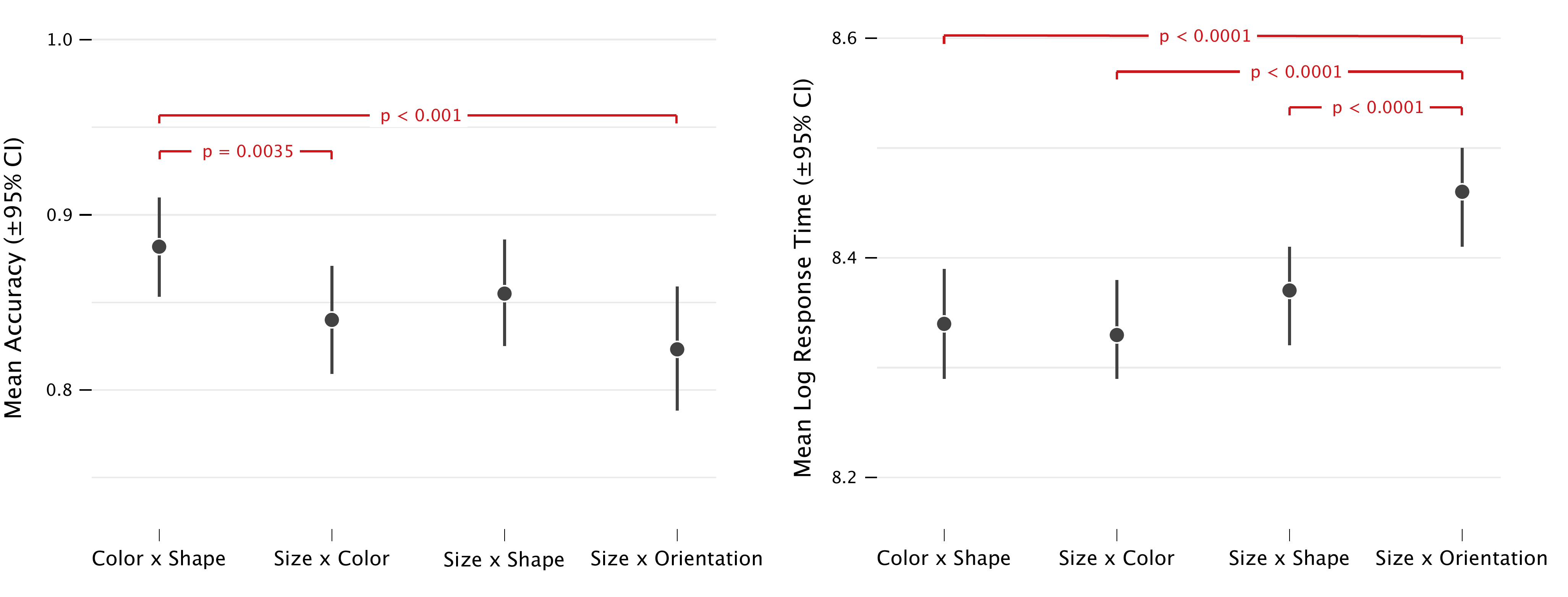}
  \Description{Overall Accuracy and Log Response Time by Channel Pair. This figure contains two plots. The left plot shows mean accuracy (with 95\% confidence intervals) for the four visual channel pairs: Color × Shape (highest accuracy), Size × Color, Size × Shape, and Size × Orientation (lowest). The right plot shows mean log response times (95\% CIs), where Size x Color is fastest and Size × Orientation is slowest. Significant pairwise differences are marked with asterisks. The accuracy axis does not start at zero; the log RT axis is truncated to the observed range.}
  \caption{Mean accuracy with 95\% CIs (left) and mean log response time with 95\% CIs (right) for the four visual channel pairs. Significant pairwise contrasts (Tukey-adjusted) are annotated. Note: Both the accuracy axis and the log response time axis do not start at zero. Both axes are shown over the observed range to highlight meaningful variation in the data.  }
  \label{means_overall}
\end{figure*}

\subsection{Overall Separability Order of Visual Channel Pairs (H1)} \hfill 

\noindent \textit{\textbf{Accuracy:}} To test H1, we grouped the data by the four visual channel pairs. 
For each participant, we computed mean accuracy for each channel pair. A repeated-measures ANOVA (Type III) revealed a significant main effect of channel pair, $F(3, 582) = 6.65, p<.001$. Mauchly’s test indicated a violation of sphericity, so we report the Greenhouse–Geisser–corrected result ($p=.00029$), confirming that accuracy differed systematically across the four channel pairs.

Mean accuracies in decreasing order were (see Figure \ref{means_overall}): 

\noindent \inlinefig{color_shape.png} = 0.882 (95\% CI [0.853, 0.910]), \inlinefig{size_shape.png} = 0.855 (95\% CI [0.825, 0.886]), \inlinefig{size_color.png} = 0.840 (95\% CI [0.809, 0.871]), and \inlinefig{size_orientation.png} = 0.823 (95\% CI [0.788, 0.859]). Tukey-adjusted pairwise comparisons showed that \inlinefig{color_shape.png} was significantly more accurate than both \inlinefig{size_color.png} ($p=.0035$) and \inlinefig{size_orientation.png} ($p=.0003$), but not different from \inlinefig{size_shape.png} ($p=.146$). No other pairwise differences were significant.

\paragraph{\textbf{Response Time:}} We calculated the mean log-transformed RT for each participant within each visual channel pair. A repeated-measures ANOVA showed a significant main effect of channel pair, $F(3,573)=18.38, p<.001$, indicating that speed also varied by visual channel pair.

Mean log RT in increasing order were (see Figure \ref{means_overall}): 

\noindent \inlinefig{size_color.png} = 8.33 (95\% CI [8.29, 8.38]), \inlinefig{color_shape.png} = 8.34 (95\% CI [8.29, 8.39], \inlinefig{size_shape.png} = 8.37 (95\% CI [8.32, 8.41]), and \inlinefig{size_orientation.png} = 8.46 (95\% CI [8.41, 8.50]). Tukey post-hoc tests revealed that \inlinefig{size_orientation.png} was significantly slower than each of the other three pairs ($p<.001$), while 

\noindent \inlinefig{color_shape.png} and \inlinefig{size_color.png} did not differ in speed ($p=.964$). 

\paragraph{\textbf{Summary:}}
\textbf{H1 is partially supported} in that both accuracy and speed analyses point to \inlinefig{color_shape.png} as the most separable (high accuracy, fast responses) and \inlinefig{size_orientation.png} as the least separable (lowest accuracy, slowest responses). However, the middle two pairs (\inlinefig{size_color.png} and \inlinefig{size_shape.png}) showed no reliable difference from one another.


\subsection{Asymmetry in Perceptual Differences (H2-H4)}

\subsubsection{Interpreting Flip Comparisons (H2–H4)} \hfill \break

\noindent H2–H4 examine performance differences between flips, that is, the two possible assignments of the task-relevant variable to one of the two channels within the same visual channel pair. However, these differences cannot be interpreted as pure asymmetries in interference between channels, because the study did not include single-channel baseline conditions. The observed differences reflect both:

\begin{itemize}
   \item  how much the non-target channel interferes with perception of the target channel, and

\item the inherent discriminability of each channel when encoding the target variable.

\end{itemize}

Thus, our flip comparisons represent overall perceptual differences between the channels in each pair, not isolated measures of interference. The observed differences indicate which channel, when assigned the target variable, yields better performance for the given pair.

\begin{figure*}[!t]
  \centering
  \includegraphics[width=0.81\textwidth]{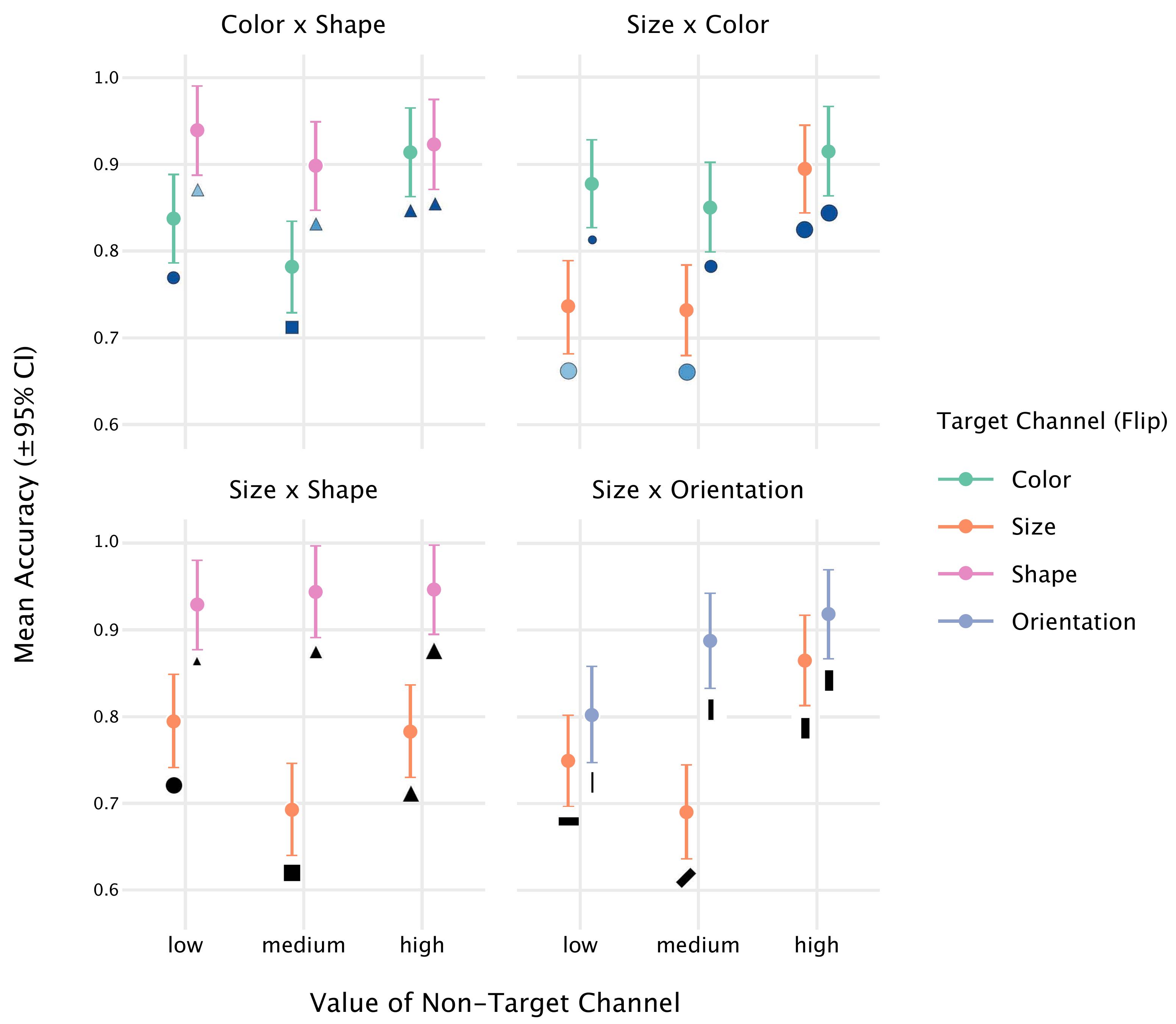}
  \Description{Accuracy by Target Channel Flip and Dimension B Value. This figure shows four panels—one per channel pair—with mean accuracy plotted for each flip (which channel encodes Dimension A). The x-axis displays the low, medium, and high values of Dimension B (the task-irrelevant dimension). Each point includes a symbol underneath depicting the specific Dimension B value (e.g., circle, square, triangle; light, medium, dark; small, medium, large; 0°, 45°, 90°). The plot shows how the target symbol’s Dimension B value influences recognition accuracy.}
  \caption{Mean accuracy with 95\% CIs for each channel pair and flip, broken down by the value of the task-irrelevant (Dimension B) attribute carried by the target symbol. Each panel represents one channel pair; within each panel, the three x-axis positions correspond to low, medium, and high levels of Dimension B. Colored points represent flips (i.e., which channel encodes the task-relevant Dimension A). To aid interpretation, the symbol beneath each point shows the actual Dimension B value paired with the target symbol in that condition. Higher accuracy indicates that the Dimension A value was easier to recognize when paired with that particular Dimension B value. Note: The accuracy axis does not start at zero and is shown over the observed range to highlight meaningful variation in the data.  }
  \label{accuracy_flips}
\end{figure*}

\begin{figure*}[!t]
  \centering
  \includegraphics[width=0.81\textwidth]{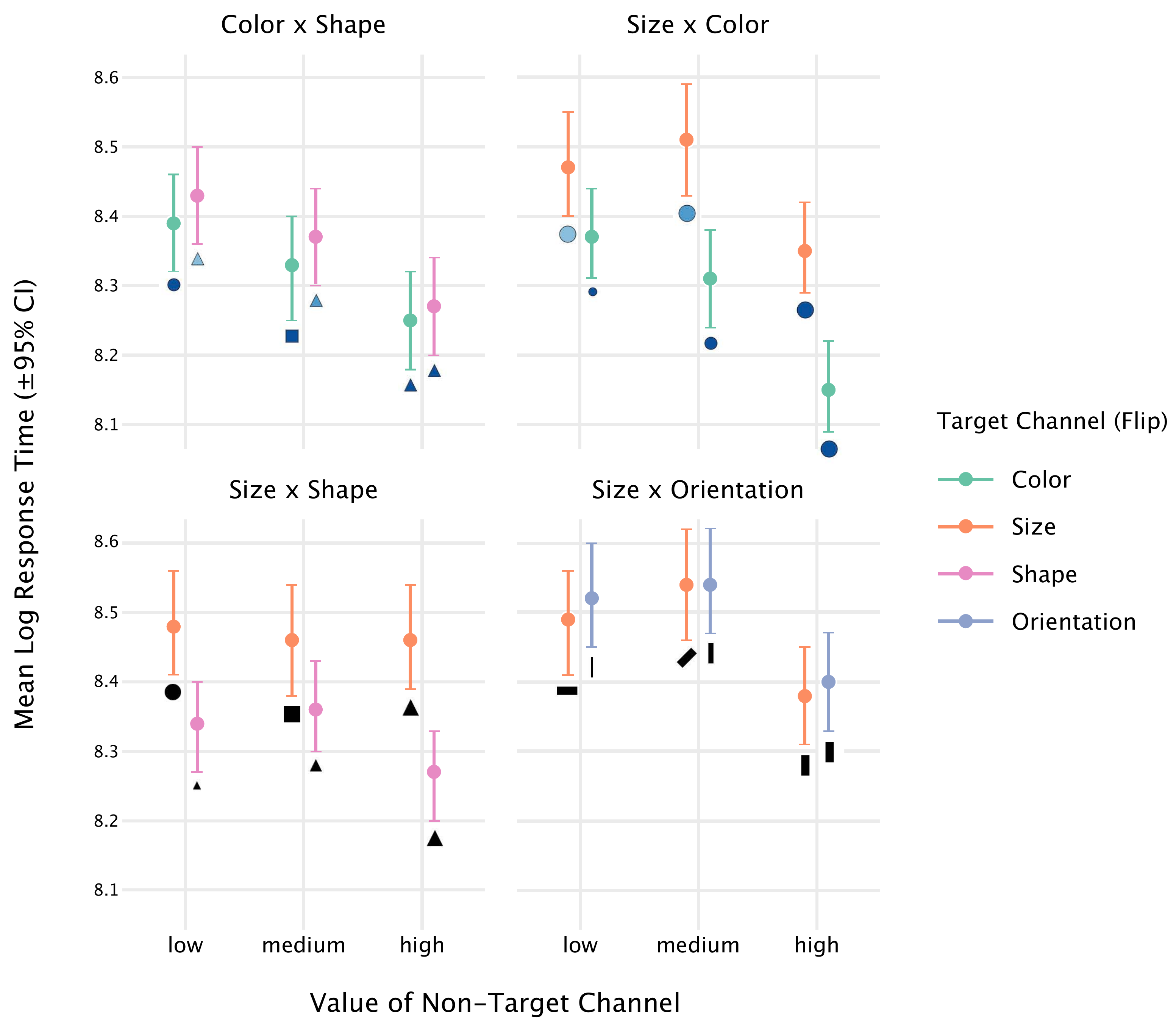}
  \Description{Log Response Time by Target Channel Flip and Dimension B Value. This figure parallels Figure 3 but plots mean log response time instead of accuracy. Four panels show the four channel pairs. Each point includes the symbol representing the Dimension B value. Lower values indicate faster responses. Across several pairs, targets with high Dimension B values (e.g., dark colors, large sizes, triangles, 90°) are identified more quickly.}
  \caption{Mean log response time with 95\% CIs for each channel pair and flip, broken down by the value of the task-irrelevant (Dimension B) attribute carried by the target symbol. Panels correspond to channel pairs, and colored points indicate the flip (task-relevant channel). As in the accuracy plots, the symbol directly beneath each mean point depicts the specific Dimension B value associated with that target symbol. Lower values on the y-axis indicate faster responses. Note: The log response time axis does not start at zero and is shown over the observed range to highlight meaningful variation in the data.  }
  \label{rt_flips}
\end{figure*}


\subsubsection{Shape x Size (H2)} \hfill 

\noindent \textit{\textbf{Accuracy:}}
We conducted a repeated-measures ANOVA on mean accuracy across the eight experimental conditions (four channel pairs × two flip assignments). Missing trial data from 20 participants were excluded listwise. Mauchly’s test indicated a violation of sphericity, $p < .001$, so Greenhouse–Geisser corrections are reported. The analysis revealed a significant main effect of condition, $F(5.45, 1232) = 16.08, p < .001, \eta_p^2 = .066$. We report the post-hoc Tukey-adjusted comparisons corresponding to H2-H4.

Mean accuracy was 0.944 (95\% CI [0.911, 0.976]) when shape encoded the target dimension and 0.760 (95\% CI [0.710, 0.810]) for size encoding the target dimension and this difference was highly significant $(p<.0001)$, indicating substantially higher accuracy when the target variable was mapped to shape (also see Figure \ref{accuracy_flips}).

\paragraph{\textbf{Response time:}}
For the RT analysis, only correctly answered trials were included, and missing trial data from 61 participants were excluded listwise. 
A repeated-measures ANOVA on mean log RT revealed a significant main effect of condition, $F(6.44, 938) = 16.53, p < .001, \eta_p^2 = .102$. Mauchly’s test indicated a modest violation of sphericity, $p = .045$, and Greenhouse–Geisser corrections are reported. We report the post-hoc Tukey-adjusted comparisons corresponding to H2-H4. 

Mean log RT was 8.278 (95\% CI [8.219, 8.337]) when shape encoded the target dimension and 8.420 (95\% CI [8.352, 8.489]) when size encoded it; responses were significantly faster when shape encoded the target dimension $(p = 0.0009)$. 

\paragraph{\textbf{Summary:}}
\textbf{H2 is supported.} Both accuracy and RT analyses indicate that if the target dimension is encoded by shape, it leads to higher accuracy and faster responses than if the target dimension were encoded by size.


\subsubsection{Color x Shape (H3)} \hfill 

\noindent \textit{\textbf{Accuracy:}} Mean accuracy was slightly higher when shape encoded the target dimension 0.913 (95\% CI [0.879, 0.948]) than when color did 0.852 (95\% CI [0.812, 0.893]). However, this difference was not statistically significant $(p = .152)$.

\paragraph{\textbf{Response Time:}}
Mean log RT was 8.263 (95\% CI [8.201, 8.324]) when color encoded the target dimension and 8.295 (95\% CI [8.232, 8.358]) when shape encoded it; the difference was not significant $(p = 0.957)$.

\paragraph{\textbf{Summary:}}
\textbf{H3 is not supported} as neither accuracy nor RT showed a significant difference between flips. However, shape was found to have significantly better accuracy for low and medium target values (see Figure \ref{accuracy_flips} below). 


\subsubsection{Color x Size (H4a)} \hfill 

\noindent \textit{\textbf{Accuracy:}}
Mean accuracy was 0.878 (95\% CI [0.839, 0.916]) when color encoded the target dimension and 0.796 (95\% CI [0.753, 0.839]) when size encoded the target dimension, and this difference was statistically significant $(p=.013)$. 

\paragraph{\textbf{Response Time:}}

Mean log RT was 8.234 (95\% CI [8.176, 8.292]) when color encoded the target dimension and 8.405 (95\% CI [8.349, 8.460]) when size encoded it; responses were significantly faster when color encoded the target dimension $(p < 0.0001)$.

\paragraph{\textbf{Summary:}} Both accuracy and RT favored color over size, indicating that participants discriminated the target variable more accurately and more quickly when it was mapped to color.


\subsubsection{Size x Orientation (H4b)} \hfill 

\noindent \textit{\textbf{Accuracy:}}
Mean accuracy was 0.867 (95\% CI [0.824, 0.910]) when orientation encoded the target dimension and 0.779 (95\% CI [0.730, 0.827]) when size encoded the target dimension and this difference was significant $(p=.043)$.

\paragraph{\textbf{Response Time:}}
Mean log RT was 8.452 (95\% CI [8.394, 8.510]) when orientation encoded the target dimension and 8.430 (95\% CI [8.368, 8.492]) when size encoded it; the difference was not significant $(p = 0.997)$.

\paragraph{\textbf{Summary:}}
Accuracy results favored orientation, but RTs showed no clear difference. This suggests an accuracy benefit without a speed advantage.

\subsection{Drill-Down by Target Value (H5)}

To examine how performance varied with the magnitude of the target, we analyzed target-value effects at the within-flip level.
For each flip within a channel pair, we tested whether accuracy and log RT differed among low, medium, and high target levels of the interfering channel. This analysis addresses whether certain target values of the interfering channel are systematically easier or harder for a given mapping. We fit separate linear mixed-effects models for accuracy and for response time, with channel pair, flip, and target value as fixed effects and participants as random intercepts. Within each flip, estimated marginal means for the three target levels were compared using Tukey-adjusted pairwise contrasts. \\

\begin{figure*}[!h]
  \centering
    \vspace*{1.5\baselineskip}
  \includegraphics[width=1\linewidth]{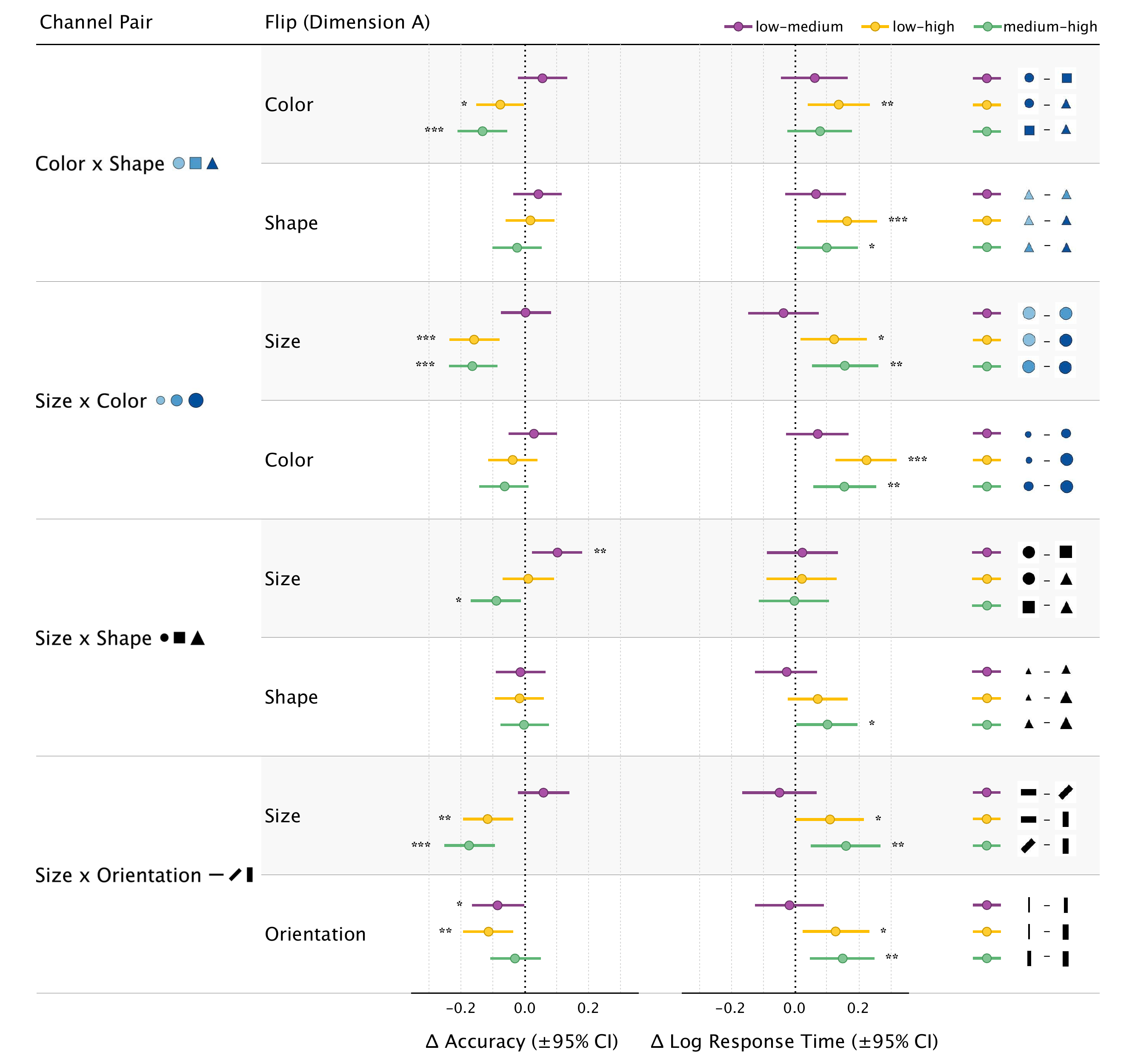}
  \vspace*{0.1\baselineskip}
  \Description{Within-Flip Effects of Dimension B on Accuracy and Response Time. This figure displays pairwise comparisons between the low–medium, low–high, and medium–high levels of Dimension B, within each flip of each channel pair. Left columns show ΔAccuracy (with 95\% CIs); right columns show ΔLog Response Time (with 95\% CIs). Positive values for accuracy indicate better performance for the first value in the pair and vice versa for log response time. Small icons next to each comparison visually depict the Dimension B values being contrasted. Statistically significant differences (Tukey-adjusted) are marked with *, **, or ***.}
\caption{Within-flip effects of Dimension B (task-irrelevant channel) on recognition of the target value of Dimension A, for each channel pair.
Each panel shows the mean accuracy and log response time differences between the low, medium, and high Dimension B values within each flip. Points represent contrast estimates (low–medium, low–high, medium–high), with horizontal bars showing 95\% CIs. The symbols in the legend to the right indicate the specific Dimension B value levels compared in each contrast. Positive $\Delta$Accuracy values 
indicate higher accuracy for the \emph{first} value in the pair (e.g., low $>$ high), 
while negative values indicate higher accuracy for the second value. For 
$\Delta$Log response time, positive values indicate \emph{slower} responses 
for the first value in the pair and \emph{faster} responses for the second; negative 
values indicate the opposite pattern. Stars denote Tukey-adjusted significance: 
$* = p < .05$, $** = p < .01$, $*** = p < .001$.}
  \label{within_flip}
  \vspace*{1\baselineskip}
\end{figure*}

\noindent \textit{\textbf{Accuracy:}}
Several flips showed reliable differences in accuracy across the levels of \emph{Dimension B} (the task-irrelevant channel), indicating that the non-target attribute of the target symbol influenced performance (see Figure~\ref{within_flip}).

For \inlinefig{color_shape.png} with \emph{color as the target channel}, accuracy was highest when the target symbol was paired with a triangle (high level of shape), compared to a circle (low) or square (medium); circle and square did not differ reliably.

For \inlinefig{size_orientation.png} with \emph{orientation as the target channel}, accuracy was lowest when the target symbol had a small size (low level of Dimension B), and higher for both medium and large sizes.

For \inlinefig{size_color.png} with \emph{size as the target channel}, accuracy was highest when the target symbol was also dark (high level of color), exceeding both the medium and light levels. Similarly, for \inlinefig{size_orientation.png} with \emph{size as the target channel}, accuracy was higher when the target symbol's orientation was 90° (high level), compared to 0° (low) or 45° (medium).

For \inlinefig{size_shape.png} with \emph{size as the target channel}, accuracy was lowest when the target symbol had the square shape (medium level of Dimension B), compared to either a circle (low) or a triangle (high).

Across flips where \emph{shape served as Dimension B}, a consistent pattern emerged: the square (medium level) was disproportionately more difficult than either the circle (low) or triangle (high), producing the largest drop in accuracy across all shape levels (see Figure~\ref{accuracy_flips}).

 \vspace*{-1\baselineskip}

\paragraph{\textbf{Response Time:}}
Response times showed a complementary pattern: the target symbol was typically identified fastest when its \emph{Dimension B} value was at the highest level (see Figure~\ref{within_flip}). This trend appeared in \inlinefig{color_shape.png} (shape encoding the target), \inlinefig{size_color.png} (both flips), and \inlinefig{size_orientation.png} (both flips).

For \inlinefig{color_shape.png} with \emph{color as the target channel}, the target symbol was identified faster when paired with a triangle (high-level shape) than when paired with a circle; circle and square did not differ reliably.

For \inlinefig{size_shape.png} with \emph{shape as the target channel}, responses were faster when the target symbol's size was large than when it was medium; the large–small difference was not reliable.

For \inlinefig{size_shape.png} with \emph{size as the target channel}, no significant RT differences were observed across the shape levels (circle, square, triangle).

\section{Discussion and Design Recommendations}

\subsection{Separability Rankings}

Our findings suggest that when designing bivariate symbol maps, pairing color with shape offers the clearest perceptual separation, producing both high accuracy and fast responses. In contrast, size with orientation was consistently the weakest combination, yielding the lowest accuracy and slowest responses, and should be avoided when rapid and reliable interpretation is important. Between the two mid-ranking pairs, \inlinefig{size_shape.png} produced slightly higher accuracy than \inlinefig{size_color.png}, but \inlinefig{size_color.png} was marginally faster. This suggests a potential trade-off: for tasks where accuracy is important, \inlinefig{size_shape.png} may be preferable, whereas for tasks where speed is more critical, \inlinefig{size_color.png} might offer an advantage. 

When compared to the frequently cited integral-separability continuum \cite [Ch. 5]{ware2019information}, our results partly align but also reveal key divergences. For example, the continuum positions \inlinefig{color_shape.png} among the more separable pairs and \inlinefig{size_orientation.png} among the less separable—patterns our study replicates. However, the mid-tier combinations do not fall into a strict linear order. Instead, \inlinefig{size_shape.png} and \inlinefig{size_color.png} formed a cluster, with their relative standing depending on whether accuracy or speed was considered. This highlights that separability may not be adequately captured by a single global ranking; rather, it can vary with the performance metric and the task context. In practice, visualization designers may need to weigh both accuracy and speed when choosing between pairings, recognizing that separability is context- and measure-dependent, rather than a fixed linear ordering.

\subsection{Asymmetries Between Flips}

Asymmetries between flips revealed that not all channels contribute equally when encoding the task-relevant variable. For example, flipping between color and shape did not affect speed, but assigning the critical variable to shape produced small but reliable gains in accuracy for low and medium values. For both \inlinefig{size_shape.png} and \inlinefig{size_color.png}, mapping the critical variable to size consistently reduced performance. 


Although our experiment did not include single-channel baseline conditions, single-channel effectiveness rankings established by prior graphical perception studies \cite{cleveland1984graphical, heer2010crowdsourcing} offer a useful reference point for understanding inherent discriminability. These studies consistently rank size to be more effective than (ordered) color and shape is ranked among the least effective categorical encodings when judged in isolation. 
Considering these rankings, our flip results reveal noteworthy reversals: size performs worse than color or shape when combined with a second channel, and shape outperforms color. These deviations from expected single-channel rankings suggest that the poorer performance of size—and the stronger performance of shape—arise from interference introduced by the paired channel, rather than solely from inherent discriminability. In other words, size appears more susceptible to cross-channel interference in bivariate symbol maps, whereas shape is comparatively robust when used as the target channel. Prior work by Smart and Szafir \cite{smart2019measuring} similarly showed that shape can strongly affect the perception of both color and size, while the reverse effects tend to be weaker. 
Together, these results suggest that effectiveness rankings derived from isolated channels do not necessarily transfer to multivariate contexts.

Our findings also suggest that not all performance benefits come with trade-offs: in some cases, stronger channels improved both speed and accuracy simultaneously. For instance, in \inlinefig{size_shape.png}, assigning the critical variable to shape improved both measures; in \inlinefig{size_color.png}, mapping to color likewise boosted both; and in \inlinefig{size_orientation.png}, orientation improved accuracy without slowing responses.

\subsection{Interpreting Value-Level Effects Within Flips}

The within-flip analyses indicate that certain values of \emph{Dimension B} made the target value of \emph{Dimension A} easier or harder to recognize. Although Dimension B was irrelevant to the task, its specific value influenced how readily the target symbol stood out from distractors. Intuitively, one might expect perceptually extreme values of the irrelevant dimension—such as darker colors or larger sizes—to make the combined A–B symbol more distinctive and therefore easier to identify. This expectation was largely borne out for color and size: targets paired with dark colors or large sizes were identified more accurately than those paired with lighter or mid-sized values.

Shape, however, deviated from this pattern. Prior work suggests that squares are often more perceptually salient or distinctive than circles or triangles \cite{demiralp2014learning, Qi2022}, yet in our data, squares were associated with the \emph{lowest} accuracy when serving as Dimension B. Orientation showed a related asymmetry: targets paired with 90° orientations were recognized more easily, while those paired with mid-level orientations (e.g., 45°) tended to yield reduced accuracy.

Taken together, these findings suggest that the recognizability of the target value depends not only on which channels are paired but also on the \emph{specific} value of the irrelevant dimension. Perceptually extreme or distinctive Dimension B values (large, dark, triangle, 90°) tended to produce more identifiable A–B combinations, whereas mid-range or low values (square, low or mid size, color, orientation) often reduced recognition. Designers should therefore consider how particular value choices within Dimension B may inadvertently sharpen or obscure the visual signature of the target encoding, even when the overall channel pairing is nominally separable.

\begin{table*}[t]
\centering
  \includegraphics[width=1\textwidth]{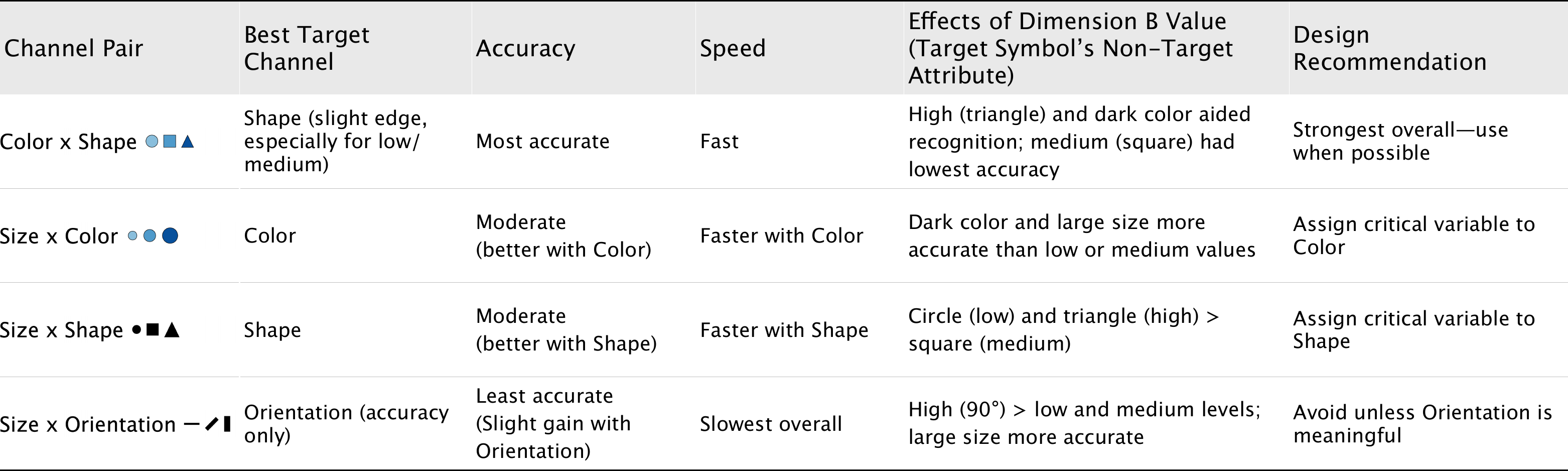}
  \Description{Summary of Design Recommendations. This table summarizes recommended mappings for bivariate symbol maps. Columns list: Channel Pair, Best Target Channel, Accuracy characteristics, Speed characteristics, Effects of Dimension B value, and Design Recommendation. Key findings include: Color × Shape is the strongest pairing; Size × Orientation is weakest; size as a target is often disadvantaged; extreme Dimension B values (dark, large, triangle, 90°) improve recognizability; squares as Dimension B reduce accuracy.}
   \caption{Summary of design guidelines for bivariate symbol maps derived from our experiment.}
   \label{design_guideline_chart}
\end{table*}

\subsection{Design Recommendations}

Based on our findings, we recommend the following guidelines for bivariate symbol map design (also see Table \ref{design_guideline_chart}):

\begin{enumerate}
    \item \textit{Prioritize color and shape for critical variables.}
    \begin{itemize}
        \item Use \inlinefig{color_shape.png} encodings when possible; they offer the clearest separation and robust performance.
        \item Either channel can encode the task-relevant variable, though assigning it to shape confers small advantages at low and medium values.
    \end{itemize}

    \item \textit{Use size cautiously, especially as the target channel.}

\begin{itemize}
        \item In \inlinefig{size_color.png}, map the critical variable to color for both higher accuracy and faster responses.
        \item In \inlinefig{size_shape.png}, map the critical variable to shape, which was substantially more accurate and faster than size.
    \end{itemize}

\item \textit{Leverage orientation selectively.}

\begin{itemize}
        \item In \inlinefig{size_orientation.png}, mapping the target to orientation improved accuracy but not speed. This pair may be useful where orientation has strong semantic relevance.
        
    \end{itemize}

\item \textit{Account for the value levels of the non-target (interfering) channel.}

\begin{itemize}
        \item High values of the interfering channel (e.g., dark color, large size, triangle, 90°) generally made the target symbol easier to recognize. When applications emphasize detecting extreme values, these high-level combinations can provide more robust discriminability.

        \item Certain mid-level values of the interfering channel—most notably the square shape—were disproportionately difficult to discriminate. If accurate recognition across all value levels is important, designers should avoid relying heavily on mid-range values of the non-target channel or consider adjusting symbol vocabularies to improve distinctiveness.
        
    \end{itemize}
    
\item \textit{Consider pairing and flip together.}
\begin{itemize}
        \item Channel pairing alone is not sufficient; the assignment of the target variable (the flip) can determine whether accuracy and speed advantages are realized. Designers should explicitly test candidate mappings rather than assuming symmetric performance.
        
    \end{itemize}

\end{enumerate}

\section{Limitations and Future Directions}



Our study, like other controlled experiments, has limitations that highlight directions for future work. Our design focused on symbol maps with one symbol per state and three discrete levels per channel. While this simplified design supports controlled comparison, real-world symbol maps often contain overlapping symbols, larger sets of value levels, and more heterogeneous spatial arrangements. While our stimuli used a fixed map layout to isolate perceptual separability from other cartographic variables, future work could systematically vary map density, marker spacing, or background rendering styles to examine how separability behaves in more complex symbol-map contexts. Future work should also examine how these more realistic conditions, particularly overlap and density, influence interference and separability.

While we tested only four channel pairs, expanding to include other channels, such as, color hue and motion, in map-based contexts would provide a broader empirical foundation. Relatedly, our study did not investigate whether integral dimensions redundantly encoding the same attribute might improve performance, a classic rationale for integral encodings \cite[Ch. 5]{ware2009quantitative}. Testing redundant and conflicting channel pairings would help clarify the balance between perceptual gain and interference.

Our task was limited to identification—finding the state with the highest value on a target dimension. While sufficient for probing separability, other tasks such as ranking, trend detection, or correlation assessment may draw differently on perceptual mechanisms and reveal distinct interference patterns.
Prior work surveying perception-based visualization studies has shown that the choice of task strongly shapes what aspects of perception are measured and which biases or limitations are uncovered \cite{quadri2021survey}. 

Our choice of symbol levels also introduced constraints. The shapes we used may be considered a relatively coarse ordinal set. As Ware’s concept of Quantitative Texton Sequences (QTonS) illustrates \cite{ware2009quantitative}, shapes can be systematically designed to convey ordered data, suggesting a direction for refining bivariate symbol design. 
In our study, all channels were discretized into three ordinal levels to ensure comparability across shape, size, and color. While shape is effective for small ordinal sets such as the three-level design used here, we note that this may not generalize to larger category sets. Nonetheless, this design reflects many real-world multivariate symbol maps, where a small number of discrete categories are often encoded with shapes while a second variable is represented using color or size in a discretized form derived from an underlying continuous metric. We also found that the square shape, when used as the non-target symbol attribute, was disproportionately difficult to discriminate—an outcome that runs counter to common assumptions about its perceptual distinctiveness. Future work should investigate why square appears to hinder recognition in this context, and whether this effect generalizes across tasks, layouts, and symbol vocabularies.

A key methodological limitation is that our study design did not include single-channel baseline conditions, which are often used to measure inherent discriminability independently of interference. Our fully bivariate design necessarily blends these two sources of performance differences. Consequently, observed asymmetries should be interpreted as overall perceptual advantages rather than pure interference effects. Future work can incorporate isolated single-channel tasks to more cleanly decompose these components.

Finally, like prior crowdsourced studies of separability \cite{smart2019measuring}, our online setting trades experimental control for ecological validity. Hardware variability may have influenced stimulus rendering across participants, though large sample sizes help mitigate these effects in aggregate. Future studies could combine controlled laboratory settings with crowdsourcing to balance precision and generalizability.

More broadly, our results show that separability is not fully captured by a single linear continuum. Accuracy and response time did not always align, and some channel pairs clustered rather than forming a strict rank order. This points to the need for models that move beyond static continua toward richer representations of separability—ones that incorporate multiple performance metrics, task demands, and contextual factors. Such models would enhance the predictive power of the integral–separable framework and provide more nuanced guidance for visualization design.

\section{Conclusion}

This study advances the empirical understanding of how visual channels interact in map-based visualizations by systematically testing the separability of four channel pairs in bivariate symbol maps. Our results show that \inlinefig{color_shape.png} offers the clearest perceptual separation, whereas \inlinefig{size_orientation.png} consistently produces the weakest performance. The two mid-ranking pairs, \inlinefig{size_shape.png} and \inlinefig{size_color.png}, formed a cluster rather than a strict order, with their relative advantage depending on whether accuracy or speed was emphasized. Beyond pairwise rankings, we observed pronounced asymmetries: assigning the task-relevant variable to color or shape improved both accuracy and speed, while size was consistently disadvantaged and orientation conferred accuracy gains without speed costs. Target values of the interfering channel also mattered—high levels were easiest to detect, while the square shape proved unexpectedly difficult, underscoring how individual values influence recognizability.

Together, these findings demonstrate that separability is not a fixed property of visual channels but depends on channel assignment, task, and value distribution of the interfering channel. They also show that effectiveness rankings derived from isolated channels do not directly transfer to multivariate contexts, where interference between channels can reshape performance. 
Ultimately, our results provide actionable guidance for designing bivariate symbol maps and contribute to refining the theoretical framework of the integral–separable continuum toward models that integrate accuracy, efficiency, and context.

\begin{acks}
We are very thankful to the reviewers for
providing valuable feedback which helped us make substantial improvements to the paper. 
This work was supported by the National Science Foundation under
Grant CMMI-1953135.
\end{acks}

\bibliographystyle{ACM-Reference-Format}
\bibliography{sample-base}


\begin{thebibliography}{30}


\ifx \showCODEN    \undefined \def \showCODEN     #1{\unskip}     \fi
\ifx \showISBNx    \undefined \def \showISBNx     #1{\unskip}     \fi
\ifx \showISBNxiii \undefined \def \showISBNxiii  #1{\unskip}     \fi
\ifx \showISSN     \undefined \def \showISSN      #1{\unskip}     \fi
\ifx \showLCCN     \undefined \def \showLCCN      #1{\unskip}     \fi
\ifx \shownote     \undefined \def \shownote      #1{#1}          \fi
\ifx \showarticletitle \undefined \def \showarticletitle #1{#1}   \fi
\ifx \showURL      \undefined \def \showURL       {\relax}        \fi
\providecommand\bibfield[2]{#2}
\providecommand\bibinfo[2]{#2}
\providecommand\natexlab[1]{#1}
\providecommand\showeprint[2][]{arXiv:#2}

\bibitem[Brainard and Wandell(1992)]%
        {brainard1992asymmetric}
\bibfield{author}{\bibinfo{person}{David~H. Brainard} {and} \bibinfo{person}{Brian~A. Wandell}.} \bibinfo{year}{1992}\natexlab{}.
\newblock \showarticletitle{Asymmetric color matching: How color appearance depends on the illuminant}.
\newblock \bibinfo{journal}{\emph{Journal of the Optical Society of America A}} \bibinfo{volume}{9}, \bibinfo{number}{9} (\bibinfo{year}{1992}), \bibinfo{pages}{1433--1448}.
\newblock
\href{https://doi.org/10.1364/JOSAA.9.001433}{doi:\nolinkurl{10.1364/JOSAA.9.001433}}


\bibitem[Burlinson et~al\mbox{.}(2017)]%
        {burlinson2017open}
\bibfield{author}{\bibinfo{person}{David Burlinson}, \bibinfo{person}{Kalpathi Subramanian}, {and} \bibinfo{person}{Paula Goolkasian}.} \bibinfo{year}{2017}\natexlab{}.
\newblock \showarticletitle{{Open vs. Closed Shapes: New Perceptual Categories?}}
\newblock \bibinfo{journal}{\emph{IEEE transactions on visualization and computer graphics}} \bibinfo{volume}{24}, \bibinfo{number}{1} (\bibinfo{year}{2017}), \bibinfo{pages}{574--583}.
\newblock
\href{https://doi.org/10.1109/TVCG.2017.2745086}{doi:\nolinkurl{10.1109/TVCG.2017.2745086}}


\bibitem[Cleveland and McGill(1984)]%
        {cleveland1984graphical}
\bibfield{author}{\bibinfo{person}{William~S Cleveland} {and} \bibinfo{person}{Robert McGill}.} \bibinfo{year}{1984}\natexlab{}.
\newblock \showarticletitle{{Graphical Perception: Theory, Experimentation, and Application to the Development of Graphical Methods}}.
\newblock \bibinfo{journal}{\emph{Journal of the American statistical association}} \bibinfo{volume}{79}, \bibinfo{number}{387} (\bibinfo{year}{1984}), \bibinfo{pages}{531--554}.
\newblock
\href{https://doi.org/10.1080/01621459.1984.10478080}{doi:\nolinkurl{10.1080/01621459.1984.10478080}}


\bibitem[Demiralp et~al\mbox{.}(2014)]%
        {demiralp2014learning}
\bibfield{author}{\bibinfo{person}{{\c{C}}a{\u{g}}atay Demiralp}, \bibinfo{person}{Michael~S Bernstein}, {and} \bibinfo{person}{Jeffrey Heer}.} \bibinfo{year}{2014}\natexlab{}.
\newblock \showarticletitle{{Learning Perceptual Kernels for Visualization Design}}.
\newblock \bibinfo{journal}{\emph{IEEE transactions on visualization and computer graphics}} \bibinfo{volume}{20}, \bibinfo{number}{12} (\bibinfo{year}{2014}), \bibinfo{pages}{1933--1942}.
\newblock
\href{https://doi.org/10.1109/TVCG.2014.2346978}{doi:\nolinkurl{10.1109/TVCG.2014.2346978}}


\bibitem[Dobson(1983)]%
        {dobson1983visual}
\bibfield{author}{\bibinfo{person}{Michael~W Dobson}.} \bibinfo{year}{1983}\natexlab{}.
\newblock \showarticletitle{{Visual Information Processing and Cartographic Communication: The Utility of Redundant Stimulus Dimensions}}.
\newblock \bibinfo{journal}{\emph{Graphic communication and design in contemporary cartography}}  \bibinfo{volume}{2} (\bibinfo{year}{1983}), \bibinfo{pages}{149--175}.
\newblock


\bibitem[Garner(1974)]%
        {garner1974processing}
\bibfield{author}{\bibinfo{person}{Wendell~R Garner}.} \bibinfo{year}{1974}\natexlab{}.
\newblock \bibinfo{booktitle}{\emph{{The Processing of Information and Structure}}}.
\newblock \bibinfo{publisher}{Psychology Press}.
\newblock
\href{https://doi.org/10.4324/9781315802862}{doi:\nolinkurl{10.4324/9781315802862}}


\bibitem[Harrower and Brewer(2003)]%
        {harrower2003colorbrewer}
\bibfield{author}{\bibinfo{person}{Mark Harrower} {and} \bibinfo{person}{Cynthia~A Brewer}.} \bibinfo{year}{2003}\natexlab{}.
\newblock \showarticletitle{{Colorbrewer. Org: An Online Tool for Selecting Colour Schemes for Maps}}.
\newblock \bibinfo{journal}{\emph{The Cartographic Journal}} \bibinfo{volume}{40}, \bibinfo{number}{1} (\bibinfo{year}{2003}), \bibinfo{pages}{27--37}.
\newblock
\href{https://doi.org/10.1179/000870403235002042}{doi:\nolinkurl{10.1179/000870403235002042}}


\bibitem[Healey and Enns(2011)]%
        {healey2011attention}
\bibfield{author}{\bibinfo{person}{Christopher Healey} {and} \bibinfo{person}{James Enns}.} \bibinfo{year}{2011}\natexlab{}.
\newblock \showarticletitle{{Attention and Visual Memory in Visualization and Computer Graphics}}.
\newblock \bibinfo{journal}{\emph{IEEE transactions on visualization and computer graphics}} \bibinfo{volume}{18}, \bibinfo{number}{7} (\bibinfo{year}{2011}), \bibinfo{pages}{1170--1188}.
\newblock
\href{https://doi.org/10.1109/TVCG.2011.127}{doi:\nolinkurl{10.1109/TVCG.2011.127}}


\bibitem[Heer and Bostock(2010)]%
        {heer2010crowdsourcing}
\bibfield{author}{\bibinfo{person}{Jeffrey Heer} {and} \bibinfo{person}{Michael Bostock}.} \bibinfo{year}{2010}\natexlab{}.
\newblock \showarticletitle{{Crowdsourcing Graphical Perception: Using Mechanical Turk to Assess Visualization Design}}. In \bibinfo{booktitle}{\emph{Proceedings of the SIGCHI conference on human factors in computing systems}}. \bibinfo{pages}{203--212}.
\newblock
\href{https://doi.org/10.1145/1753326.1753357}{doi:\nolinkurl{10.1145/1753326.1753357}}


\bibitem[Kim and Heer(2018)]%
        {kim2018assessing}
\bibfield{author}{\bibinfo{person}{Younghoon Kim} {and} \bibinfo{person}{Jeffrey Heer}.} \bibinfo{year}{2018}\natexlab{}.
\newblock \showarticletitle{{Assessing Effects of Task and Data Distribution on the Effectiveness of Visual Encodings}}. In \bibinfo{booktitle}{\emph{Computer Graphics Forum}}, Vol.~\bibinfo{volume}{37}. Wiley Online Library, \bibinfo{pages}{157--167}.
\newblock
\href{https://doi.org/10.1111/cgf.13409}{doi:\nolinkurl{10.1111/cgf.13409}}


\bibitem[Liu et~al\mbox{.}(2021)]%
        {Liu_orientation}
\bibfield{author}{\bibinfo{person}{Tingting Liu}, \bibinfo{person}{Xiaotong Li}, \bibinfo{person}{Chen Bao}, \bibinfo{person}{Michael Correll}, \bibinfo{person}{Changehe Tu}, \bibinfo{person}{Oliver Deussen}, {and} \bibinfo{person}{Yunhai Wang}.} \bibinfo{year}{2021}\natexlab{}.
\newblock \showarticletitle{Data-Driven Mark Orientation for Trend Estimation in Scatterplots}. In \bibinfo{booktitle}{\emph{Proceedings of the 2021 CHI Conference on Human Factors in Computing Systems}} (Yokohama, Japan) \emph{(\bibinfo{series}{CHI '21})}. \bibinfo{publisher}{Association for Computing Machinery}, \bibinfo{address}{New York, NY, USA}, Article \bibinfo{articleno}{473}, \bibinfo{numpages}{16}~pages.
\newblock
\showISBNx{9781450380966}
\href{https://doi.org/10.1145/3411764.3445751}{doi:\nolinkurl{10.1145/3411764.3445751}}


\bibitem[Lu et~al\mbox{.}(2023)]%
        {Lu2023}
\bibfield{author}{\bibinfo{person}{Kecheng Lu}, \bibinfo{person}{Khairi Reda}, \bibinfo{person}{Oliver Deussen}, {and} \bibinfo{person}{Yunhai Wang}.} \bibinfo{year}{2023}\natexlab{}.
\newblock \showarticletitle{{Interactive Context-Preserving Color Highlighting for Multiclass Scatterplots}}. In \bibinfo{booktitle}{\emph{Proceedings of the 2023 CHI Conference on Human Factors in Computing Systems}} (Hamburg, Germany) \emph{(\bibinfo{series}{CHI '23})}. \bibinfo{publisher}{Association for Computing Machinery}, \bibinfo{address}{New York, NY, USA}, Article \bibinfo{articleno}{823}, \bibinfo{numpages}{15}~pages.
\newblock
\showISBNx{9781450394215}
\href{https://doi.org/10.1145/3544548.3580734}{doi:\nolinkurl{10.1145/3544548.3580734}}


\bibitem[Lu et~al\mbox{.}(2021)]%
        {lu2021modeling}
\bibfield{author}{\bibinfo{person}{Min Lu}, \bibinfo{person}{Joel Lanir}, \bibinfo{person}{Chufeng Wang}, \bibinfo{person}{Yucong Yao}, \bibinfo{person}{Wen Zhang}, \bibinfo{person}{Oliver Deussen}, {and} \bibinfo{person}{Hui Huang}.} \bibinfo{year}{2021}\natexlab{}.
\newblock \showarticletitle{{Modeling Just Noticeable Differences in Charts}}.
\newblock \bibinfo{journal}{\emph{IEEE transactions on visualization and computer graphics}} \bibinfo{volume}{28}, \bibinfo{number}{1} (\bibinfo{year}{2021}), \bibinfo{pages}{718--726}.
\newblock
\href{https://doi.org/10.1109/TVCG.2021.3114874}{doi:\nolinkurl{10.1109/TVCG.2021.3114874}}


\bibitem[MacEachren(2004)]%
        {maceachren2004maps}
\bibfield{author}{\bibinfo{person}{Alan~M MacEachren}.} \bibinfo{year}{2004}\natexlab{}.
\newblock \bibinfo{booktitle}{\emph{{How Maps Work: Representation, Visualization, and Design}}}.
\newblock


\bibitem[Munzner(2025)]%
        {munzner2025visualization}
\bibfield{author}{\bibinfo{person}{Tamara Munzner}.} \bibinfo{year}{2025}\natexlab{}.
\newblock \showarticletitle{{Visualization Analysis and Design}}. In \bibinfo{booktitle}{\emph{Proceedings of the Special Interest Group on Computer Graphics and Interactive Techniques Conference Courses}}. \bibinfo{pages}{1--2}.
\newblock
\href{https://doi.org/10.1145/3721241.3733989}{doi:\nolinkurl{10.1145/3721241.3733989}}


\bibitem[Oicherman et~al\mbox{.}(2008)]%
        {oicherman2008effect}
\bibfield{author}{\bibinfo{person}{Boris Oicherman}, \bibinfo{person}{Ming~Ronnier Luo}, \bibinfo{person}{Bryan Rigg}, {and} \bibinfo{person}{Alan~R. Robertson}.} \bibinfo{year}{2008}\natexlab{}.
\newblock \showarticletitle{Effect of observer metamerism on colour matching of display and surface colours}.
\newblock \bibinfo{journal}{\emph{Color Research \& Application}} \bibinfo{volume}{33}, \bibinfo{number}{5} (\bibinfo{year}{2008}), \bibinfo{pages}{346--359}.
\newblock
\href{https://doi.org/10.1002/col.20429}{doi:\nolinkurl{10.1002/col.20429}}


\bibitem[Pomerantz and Garner(1973)]%
        {pomerantz1973stimules}
\bibfield{author}{\bibinfo{person}{James~R Pomerantz} {and} \bibinfo{person}{WR Garner}.} \bibinfo{year}{1973}\natexlab{}.
\newblock \showarticletitle{{Stimules Configuration in Selective Attention Tasks}}.
\newblock \bibinfo{journal}{\emph{Perception \& Psychophysics}} \bibinfo{volume}{14}, \bibinfo{number}{3} (\bibinfo{year}{1973}), \bibinfo{pages}{565--569}.
\newblock


\bibitem[Qi and Jing(2022)]%
        {Qi2022}
\bibfield{author}{\bibinfo{person}{Guo Qi} {and} \bibinfo{person}{Zhang Jing}.} \bibinfo{year}{2022}\natexlab{}.
\newblock \showarticletitle{{The Quantitative Research on Length and Area Perception: A Guidance on Shape Encoding in Visual Interface}}.
\newblock \bibinfo{journal}{\emph{Displays}}  \bibinfo{volume}{75} (\bibinfo{year}{2022}), \bibinfo{pages}{102325}.
\newblock
\showISSN{0141-9382}
\href{https://doi.org/10.1016/j.displa.2022.102325}{doi:\nolinkurl{10.1016/j.displa.2022.102325}}


\bibitem[Quadri and Rosen(2021)]%
        {quadri2021survey}
\bibfield{author}{\bibinfo{person}{Ghulam~Jilani Quadri} {and} \bibinfo{person}{Paul Rosen}.} \bibinfo{year}{2021}\natexlab{}.
\newblock \showarticletitle{{A Survey of Perception-Based Visualization Studies by Task}}.
\newblock \bibinfo{journal}{\emph{IEEE transactions on visualization and computer graphics}} \bibinfo{volume}{28}, \bibinfo{number}{12} (\bibinfo{year}{2021}), \bibinfo{pages}{5026--5048}.
\newblock
\href{https://doi.org/10.1109/TVCG.2021.3098240}{doi:\nolinkurl{10.1109/TVCG.2021.3098240}}


\bibitem[Shortridge(1982)]%
        {shortridge1982stimulus}
\bibfield{author}{\bibinfo{person}{Barbara~G Shortridge}.} \bibinfo{year}{1982}\natexlab{}.
\newblock \showarticletitle{{Stimulus Processing Models From Psychology: Can We Use Them in Cartography?}}
\newblock \bibinfo{journal}{\emph{The American Cartographer}} \bibinfo{volume}{9}, \bibinfo{number}{2} (\bibinfo{year}{1982}), \bibinfo{pages}{155--167}.
\newblock
\href{https://doi.org/10.1559/152304082783948501}{doi:\nolinkurl{10.1559/152304082783948501}}


\bibitem[Smart and Szafir(2019)]%
        {smart2019measuring}
\bibfield{author}{\bibinfo{person}{Stephen Smart} {and} \bibinfo{person}{Danielle~Albers Szafir}.} \bibinfo{year}{2019}\natexlab{}.
\newblock \showarticletitle{{Measuring the Separability of Shape, Size, and Color in Scatterplots}}. In \bibinfo{booktitle}{\emph{Proceedings of the 2019 CHI Conference on Human Factors in Computing Systems}}. \bibinfo{pages}{1--14}.
\newblock
\href{https://doi.org/10.1145/3290605.3300899}{doi:\nolinkurl{10.1145/3290605.3300899}}


\bibitem[Stevens(2017)]%
        {stevens2017psychophysics}
\bibfield{author}{\bibinfo{person}{Stanley~Smith Stevens}.} \bibinfo{year}{2017}\natexlab{}.
\newblock \bibinfo{booktitle}{\emph{{Psychophysics: Introduction to Its Perceptual, Neural and Social Prospects}}}.
\newblock \bibinfo{publisher}{Routledge}.
\newblock
\href{https://doi.org/10.4324/9781315127675}{doi:\nolinkurl{10.4324/9781315127675}}


\bibitem[Stevens and Guirao(1963)]%
        {stevens1963subjective}
\bibfield{author}{\bibinfo{person}{Stanley~Smith Stevens} {and} \bibinfo{person}{Miguelina Guirao}.} \bibinfo{year}{1963}\natexlab{}.
\newblock \showarticletitle{{Subjective Scaling of Length and Area and the Matching of Length to Loudness and Brightness.}}
\newblock \bibinfo{journal}{\emph{Journal of Experimental Psychology}} \bibinfo{volume}{66}, \bibinfo{number}{2} (\bibinfo{year}{1963}), \bibinfo{pages}{177}.
\newblock
\href{https://doi.org/10.1037/h0044984}{doi:\nolinkurl{10.1037/h0044984}}


\bibitem[Stone et~al\mbox{.}(2014)]%
        {stone2014engineering}
\bibfield{author}{\bibinfo{person}{Maureen Stone}, \bibinfo{person}{Danielle~Albers Szafir}, {and} \bibinfo{person}{Vidya Setlur}.} \bibinfo{year}{2014}\natexlab{}.
\newblock \showarticletitle{{An Engineering Model for Color Difference as a Function of Size}}. In \bibinfo{booktitle}{\emph{Color and Imaging Conference}}, Vol.~\bibinfo{volume}{22}. Society for Imaging Science and Technology, \bibinfo{pages}{253--258}.
\newblock
\href{https://doi.org/10.2352/CIC.2014.22.1.art00045}{doi:\nolinkurl{10.2352/CIC.2014.22.1.art00045}}


\bibitem[Szafir(2017)]%
        {szafir2017modeling}
\bibfield{author}{\bibinfo{person}{Danielle~Albers Szafir}.} \bibinfo{year}{2017}\natexlab{}.
\newblock \showarticletitle{{Modeling Color Difference for Visualization Design}}.
\newblock \bibinfo{journal}{\emph{IEEE transactions on visualization and computer graphics}} \bibinfo{volume}{24}, \bibinfo{number}{1} (\bibinfo{year}{2017}), \bibinfo{pages}{392--401}.
\newblock
\href{https://doi.org/10.1109/TVCG.2017.2744359}{doi:\nolinkurl{10.1109/TVCG.2017.2744359}}


\bibitem[Tseng et~al\mbox{.}(2023)]%
        {tseng2023measuring}
\bibfield{author}{\bibinfo{person}{Chin Tseng}, \bibinfo{person}{Ghulam~Jilani Quadri}, \bibinfo{person}{Zeyu Wang}, {and} \bibinfo{person}{Danielle~Albers Szafir}.} \bibinfo{year}{2023}\natexlab{}.
\newblock \showarticletitle{{Measuring Categorical Perception in Color-Coded Scatterplots}}. In \bibinfo{booktitle}{\emph{proceedings of the 2023 CHI conference on human factors in computing systems}}. \bibinfo{pages}{1--14}.
\newblock
\href{https://doi.org/10.1145/3544548.3581416}{doi:\nolinkurl{10.1145/3544548.3581416}}


\bibitem[Tseng et~al\mbox{.}(2025)]%
        {tsengshape}
\bibfield{author}{\bibinfo{person}{Chin Tseng}, \bibinfo{person}{Arran~Zeyu Wang}, \bibinfo{person}{Ghulam~Jilani Quadri}, {and} \bibinfo{person}{Danielle~Albers Szafir}.} \bibinfo{year}{2025}\natexlab{}.
\newblock \showarticletitle{{Shape It Up: An Empirically Grounded Approach for Designing Shape Palettes}}.
\newblock \bibinfo{journal}{\emph{IEEE Transactions on Visualization and Computer Graphics}} \bibinfo{volume}{31}, \bibinfo{number}{1} (\bibinfo{year}{2025}), \bibinfo{pages}{349--359}.
\newblock
\href{https://doi.org/10.1109/TVCG.2024.3456385}{doi:\nolinkurl{10.1109/TVCG.2024.3456385}}


\bibitem[Ware(2009)]%
        {ware2009quantitative}
\bibfield{author}{\bibinfo{person}{Colin Ware}.} \bibinfo{year}{2009}\natexlab{}.
\newblock \showarticletitle{{Quantitative Texton Sequences for Legible Bivariate Maps}}.
\newblock \bibinfo{journal}{\emph{IEEE Transactions on Visualization and Computer Graphics}} \bibinfo{volume}{15}, \bibinfo{number}{6} (\bibinfo{year}{2009}), \bibinfo{pages}{1523--1530}.
\newblock
\href{https://doi.org/10.1109/TVCG.2009.175}{doi:\nolinkurl{10.1109/TVCG.2009.175}}


\bibitem[Ware(2019)]%
        {ware2019information}
\bibfield{author}{\bibinfo{person}{Colin Ware}.} \bibinfo{year}{2019}\natexlab{}.
\newblock \bibinfo{booktitle}{\emph{{Information Visualization: Perception for Design}}}.
\newblock \bibinfo{publisher}{Morgan Kaufmann}.
\newblock
\href{https://doi.org/10.5555/2285540}{doi:\nolinkurl{10.5555/2285540}}


\bibitem[Zeng and Battle(2023)]%
        {zeng2023review}
\bibfield{author}{\bibinfo{person}{Zehua Zeng} {and} \bibinfo{person}{Leilani Battle}.} \bibinfo{year}{2023}\natexlab{}.
\newblock \showarticletitle{{A Review and Collation of Graphical Perception Knowledge for Visualization Recommendation}}. In \bibinfo{booktitle}{\emph{Proceedings of the 2023 CHI conference on human factors in computing systems}}. \bibinfo{pages}{1--16}.
\newblock
\href{https://doi.org/10.1145/3544548.3581349}{doi:\nolinkurl{10.1145/3544548.3581349}}


\end{thebibliography}


\end{document}